\documentclass{article} 
\usepackage{nips12submit_e,times}

\nipsfinalcopy 
\usepackage{natbib}
\usepackage{comment}
\usepackage{graphicx} 
\usepackage{subfigure}
\usepackage{hyperref}
\usepackage{url}
\usepackage{amsmath, amsthm, amssymb, multirow, paralist}
\usepackage{algorithm,algorithmic}
\usepackage{amssymb, multirow, paralist}

\newtheorem{thm}{Theorem}
\newtheorem{lemma}{Lemma}

\usepackage{hyperref}
\hypersetup{
colorlinks=true,
citecolor=blue,
linkbordercolor={1  0 1}, 
citebordercolor={1 1 1} 
}

\def \R {\mathbb{R}}

\def \y {\mathbf{y}}

\def \E {\mathrm{E}}

\def \y {\mathbf{y}}

\def \Y {\mathcal{Y}}

\def \R {\mathbb{R}}

\def \I {\mathcal{I}}

\def \R {\mathbb{R}}

\def \y {\mathbf{y}}

\def \E {\mathrm{E}}

\def \y {\mathbf{y}}

\def \R {\mathbb{R}}

\def \I {\mathcal{I}}

\title{Analysis of Distributed Stochastic\\ Dual Coordinate Ascent}

\author{
Tianbao Yang, Shenghuo Zhu, Rong Jin$^\dagger$, Yuanqing Lin \\
NEC Labs America, Cupertino, CA 95014\\
$^\dagger$Michigan State University, East Lansing, MI 48824\\
\texttt{\{tyang, zsh, ylin\}@nec-labs.com, rongjin@cse.msu.edu}
}

\begin{document}

\maketitle
\begin{abstract} 
In~\citep{Yangnips13}, the author presented distributed stochastic dual coordinate ascent (DisDCA) algorithms for solving large-scale regularized loss minimization.  Extraordinary performances have been observed and reported for the well-motivated updates, as referred to the practical updates,  compared to the naive updates. However, no serious analysis has been provided to understand the updates and therefore the convergence rates. In the paper, we bridge the gap by providing a theoretical analysis of the convergence rates of the practical 
DisDCA algorithm. Our analysis helped by empirical studies has  shown that it could yield an exponential speed-up in the convergence by increasing the number of dual updates at each iteration. This result justifies the superior performances of the practical DisDCA as compared to the naive variant. As a byproduct, our analysis also   reveals the  convergence behavior of the one-communication DisDCA.
\end{abstract}

\section{Introduction}
With the exponential growth of data, it has become an urgent task to design distributed (or parallel) optimization for big data analytics. The surge of a large cluster of machines has made the distributed optimization possible. The goal of distributed optimization is to optimize a certain objective defined over millions of billions of data that is distributed over many machines by utilizing the computational power of these machines. 

The key concern in distributed optimization is how to coordinate the communication between so many machines such that the latency is minimized while the convergence performance is maximized. In this work, we focus on a particular distributed optimization algorithm, i.e., distributed stochastic dual coordinate ascent (DisDCA). The idea behind the stochastic dual coordinate ascent is to update the dual variables to increase the dual objective. It has been proven and observed that the stochastic dual coordinate ascent could achieve extraordinary performances in optimizing regularized loss minimization problems (e.g., SVMs, least square regression, and logistic regression). 

The mechanism of DisDCA is to introduce an sequence of $m$ updates on individual machines before performing a communication between machines. The motivation is that if the speed of gain in convergence is faster than the speed of incurred computation, the increasing $m$ would alleviate the communication demand. Two variants of DisDCA~\citep{Yangnips13}(the basic variant and the practical variant) have been proposed and compared empirically. Although the basic one (as referred to the naive variant in this paper) has been analyzed, however, its empirical performance is significantly worse than the practical variant. The contribution of the paper is to present some theoretical analysis as well as empirical studies of the practical DisDCA, which provides  more insights into the practical DisDCA. In particular, we first prove the the practical DisDCA for orthogonal data (data on different machines or orthogonal) and establish its convergence. The result is not only interesting by itself, but also relates to the one-communication distributed optimization. Moreover, we analyze the practical DisDCA for general cases. Our analysis under the help of empirical studies have shown that it could yield an \textit{exponential} speed-up in convergence by increasing $m$, which is significantly faster as compared to the partially linear speed-up of the naive variant.

\label{submission}
\section{DisDCA: the practical updates versus the naive updates}
We begin with a description of the practical updates and the naive updates of DisDCA. To this end, we introduce some notations. 

Let $(x_i, y_i), i=1,\ldots, n$ denote a set of $n$ training data with feature factor $x_i\in\R^d$ and the label $y_i\in\Y$. Assume the training data are evenly distributed over $K$ machines, and let $X_k=(x^k_1,\ldots, x^k_{n_k})$ denote the feature vectors of data on the $k$-th machine. The goal is to solve the following optimization problem: 
\begin{align}\label{eqn:primal}
\begin{aligned}
&\min_{w\in\R^d} P(w)\\
&\text{where } P(w) = \frac{1}{n}\sum_{i=1}^n\phi(x_i\cdot w, y_i) + g(w)
\end{aligned}
\end{align}
where $x\cdot w$ denotes the inner product of two vectors, $\phi(x\cdot w, y)$ denote a convex loss function w.r.t the first argument, and $g(w)$ denotes a convex regularizer. In this paper, we focus on smooth loss function $\phi(z,y)$  and strongly convex regularizer $g(w)$, which satisfying the following properties, respectively: 
\begin{align}
&|\nabla\phi(z_1) - \nabla\phi(z_2)|\leq L|z_1 - z_2|\\
&\|\nabla g(w) - \nabla g(w)\|_2\geq \lambda\|w_1-w_2\|_2
\end{align}
where $L$ characterizes the smoothness of $\phi(z)$ and $\lambda$ characterizes the strong convexity of $g(w)$. In order to solve the above problem by SDCA, we cast the primal problem in~(\ref{eqn:primal}) into a dual problem. To this end, we introduce two notations, $\phi^*(\alpha)$ and $g^*(u)$ to denote the convex conjugate of $\phi(z, y)$ and $g(w)$, respectively. Assuming $g^*(u)$ is continuous function, we can cast the problem into the following dual problem: 
\begin{align}\label{eqn:dual}
\begin{aligned}
&\max_{\alpha\in\R^n} D(\alpha)\\
&\text{where }\frac{1}{n} \sum_{i=1}^n-\phi_i^*(-\alpha_i)  - g^*\left(\frac{1}{n}\sum_{i=1}^n\alpha_ix_i\right)
\end{aligned}
\end{align}
The correspondence between the optimal solution $w^*$ to ~(\ref{eqn:primal}) and the optimal solution $\alpha^*$ to~(\ref{eqn:dual}) is given by 
\begin{align}
w^* = \nabla g^*(v^*), \quad v^* =\frac{1}{n}\sum_{i=1}^n\alpha_i x_i
\end{align}
Before proceeding, let us recall several important applications of SDCA. For classification, we can choose the squared hinge loss function $\phi(z, y) = (1 - z)_+^2$ or the logistic loss $\phi(z, y) =  \log (1 + \exp(-yz))$; for regression one can consider the least square loss $\phi(z, y) =|y-z|^2/2$. It is not difficult to  derive that squared  hinge loss is $2$-smooth convex function, logistic loss is $1/2$-smooth convex function, and least square loss is $1$-smooth convex function. In terms of regularizer, one can consider the $\ell_2$ norm square regularizer $g(w) = \frac{\lambda}{2}\|w\|_2^2$ or the elastic net regularizer $g(w) = \frac{\lambda}{2}\|w\|_2^2 + \mu\|w\|_1$, which are all $\lambda$-strongly convex regularizer.  In the following analysis, we let $c = L/\lambda$ denote the condition number of the problem, which  is an important factor that impacts the convergence of optimization. 

\begin{algorithm}[tb]
\caption{\bfseries DisDCA Algorithm}
 \label{alg:disdca}
\begin{algorithmic}
   \STATE {\bfseries Input:} iterations $T$, size $m$, size $K$
   \STATE {\bfseries Excute:} \textbf{mR-SDCA}($m, T$) on $K$ machines 
   \end{algorithmic}

\vspace*{0.1in}
\centering{\bfseries mR-SDCA} on machine $k$\vspace*{0.1in}
\noindent\fbox{
\begin{minipage}{0.95\linewidth}
\begin{algorithmic}
   \STATE {\bfseries Input:} iterations $T$, size $m$
      \STATE {\bfseries Load Data:} $X_k, \y_k$
   \STATE {\bfseries Initializations:} $\alpha^0=0$, $w^0_k=w^0=\nabla g^*(0)$
   \FOR{$t=1$ {\bfseries to} $T$}
   \STATE Sample $m$ examples randomly, indexed by ${\I_m}$
   \STATE Update dual vars by \textbf{Inc-Dual}($w^{t-1}, scale, \I_m$)
   \STATE Update $w_k^{t} = \frac{K}{\lambda n}\sum_{i=1}^{n_k}\alpha_{k,i}^tx_{k,i}$
   \STATE Reduce $w^t = \frac{1}{K}\sum_{k=1}^Kw_k^t$
   \ENDFOR
\end{algorithmic}
\end{minipage}
}
\end{algorithm}

To facilitate the understanding of the algorithm and the proof,  in the sequel we simply assume $g(w) = \lambda /2 \|w\|_2^2$, and thus $g^*(v) = \frac{1}{2\lambda}\|v\|_2^2$. A slight modification of the algorithm and a careful examination of  the analysis reveals that the results hold for any strongly convex function $g(w)$ by exploring the following inequality: 
\begin{align*}
g^*(v + \Delta v) \leq g^*(v) + \Delta v \cdot \nabla g^*(v) + \frac{1}{2\lambda}\|\Delta v\|_2^2
\end{align*}
due to that the convex conjugate $g^*(v)$ of a $\lambda$-strongly convex function $g(w)$ is a $1/\lambda$ smooth convex function.

We present in Algorithm~\ref{alg:disdca} the DisDCA algorithm. The algorithm deployes $K$ processes \textbf{mR-SDCA} on $K$ different machines that work on different subsets of data. The procedure \textbf{mR-SDCA} runs a total of $T$-iterations and at each iteration samples $m$ examples and calls \textbf{Inc-Dual} to update the dual variables of the sampled examples.  Algorithm~\ref{alg:inc} drafts the two different variants of the DisDCA for updating the sampled dual variables, where \textbf{n-variant} refers to the naive variant and \textbf{p-variant} refers to the practical variant. 
\begin{algorithm}[t]
\caption{\bfseries Inc-Dual($w^{t-1}, scale, \I_m$)}
 \label{alg:inc}
\begin{algorithmic}[H]
   \STATE {\bfseries Input:} $w^{t-1}, scale$ and $\I_m$\\
   //$scale=mK$ if {\bfseries n-variant}, otherwise  $K$
   \vspace*{0.06in}
   \STATE{\bfseries p-variant Initializations:} $ u^{t,0}_k = w^{t-1}, \alpha_k^{t,0}=\alpha_k^{t-1}$
   \FOR{$j=1$ {\bfseries to} $m$}
   \STATE Let $i=\I_m[j]$
   \STATE {\bfseries n-variant:} Compute $\Delta\alpha_{k,i}$ by solving~(\ref{eqn:dual-n})
   
   \STATE {\bfseries p-variant: }Compute $\Delta\alpha_{k,i}$ by solving~(\ref{eqn:dual-p})
   \STATE {\bfseries p-variant: }Update $u_k^{t, j} = u^{t,j-1}_k + \frac{K}{\lambda n}\Delta\alpha_{k,i}x_{k,i}$
   \STATE Update $\alpha^{t, j}_{k,i}=\alpha^{t, j-1}_{k,i} + \Delta\alpha_{k, i}$
   \ENDFOR
\end{algorithmic}
\end{algorithm}

The difference between the naive variant and the practical variant lies on how to update the dual variables at each iteration. In the naive variant, all the dual variables are updated using the same primal solution $w^{t-1}$, i.e.,  
\begin{align}\label{eqn:dual-n}
 \Delta\alpha_{k,i}= \max_{\Delta\alpha}&-\phi_{k,i}^*(-(\alpha_{k,i}^{t-1} + \Delta\alpha)) - \Delta\alpha x_{k,i}^{\top}{\color{red}w^{t-1}}\nonumber\\
 & - \frac{{\color{red}[scale=mK]}}{2\lambda n}(\Delta\alpha)^2\|x_{k,i}\|_2^2,
 \end{align}
 while the dual updates in the practical variant is achieved by using and updating local primal solution $u^{t, j}_k$, i.e.,
\begin{align}\label{eqn:dual-p}
 \Delta\alpha_{k,i}&= \max_{\Delta\alpha}-\phi_{k,i}^*(-(\alpha_{k,i}^{t-1} + \Delta\alpha)) - \Delta\alpha x_{k,i}^{\top}{\color{red}u_k^{t, j-1}}\nonumber\\
 & \hspace*{0.4in}- \frac{\color{red}[scale=K]}{2\lambda n}(\Delta\alpha)^2\|x_{k,i}\|_2^2\\
u^{t, j}_k  &= u_k^{t, j-1} + \frac{K}{\lambda n}\Delta\alpha_{k,i}x_{k,i}\nonumber
 \end{align}
Intuitively, the practical variant makes use of the updated information as in the updated dual variables and therefore in the updated local primal solution $u^{t,j}_k$, therefore it is of no surprise that it can have better performances than the naive variants. From another point of view, by utilizing the updated solution $u^{t,j}_k$, the practical DisDCA is abel to use a larger step size (corresponding to smaller $scale$) in updating $\alpha$. 

It has been shown in~\citep{Yangnips13}, the procedure in {\bf Inc-Dual} for the practical variant is to increase the objective of the following sub-dual problem: 
\begin{align}\label{eqn:sub}
&\max_{\alpha}\frac{1}{n_k} \sum_{i\in\I_m}-\phi_{k,i}^*(-\alpha_{k,i})\\
& + \frac{\lambda }{2}\left\| w^{t-1} - \frac{1}{\lambda n_k}\sum_{i\in\I_m}\alpha^{t-1}_{k,i}x_{k,i}+\frac{1}{\lambda n_k}\sum_{i\in\I_m}\alpha_{k,i}x_{k,i}\right\|_2^2\nonumber
\end{align}
by employing SDCA to update the sampled dual variables once with initial dual solutions $\alpha^{t,0}_{k,i}=\alpha^{t-1}_{k,i}$.

Figure~\ref{fig:1} shows an empirical comparison between the two variants in optimizing  SVM with squared hinge loss.  It clearly demonstrates the faster convergence of the practical DisDCA versus the naive DisDCA.   While ~\citet{Yangnips13} has established  the convergence rate of the naive DisDCA, however, it still remains an open problem and is of great interest to analyze the convergence of the practical DisDCA. In the next two sections, we provide a theoretical analysis as well as empirical studies to justify the practical DisDCA. For comparison, we state in the following theorem the convergence result of the native DisDCA.

\begin{thm}[\citet{Yangnips13}]\label{thm:1}
Assume all data points are in the unit ball, i.e.,  $\|x\|_2\leq 1$. For a $L$-smooth loss function $\phi_i$ and a $\lambda$-strongly convex function $g(w)$, let $w^T, \alpha^T$ be the primal and dual solution obtained  by the DisDCA algorithm with the naive updates in~(\ref{eqn:dual-n}), then we have
\begin{align*}
\E\left[D(\alpha^*) - D(\alpha^T)\right] \leq \left(1 - \frac{1}{c + \frac{n}{mK}}\right)^T\epsilon_0
\end{align*}
and 
\begin{align*}
\E\left[P(w^T) - D(\alpha^T)\right] \leq \left(c + \frac{n}{mK}\right)\left(1 - \frac{1}{c + \frac{n}{mK}}\right)^T\epsilon_0
\end{align*}
where $c= L/\lambda$ is the condition number and $\epsilon_0 = D(\alpha^*) - D(\alpha^0)\leq P(w^0) - D(\alpha^0)$ is a constant.
\end{thm}
{\bf Remark:} From Theorem~\ref{thm:1}, we can see that the effective region, where increasing $m$ and $K$ can improve the convergence rate, is heavily impacted by the condition number $c=L/\lambda$. In particular, when $c = \Omega(n)$, the benefit of increasing $m$ and $K$ becomes very small because the term $c + n/(mK)$ is dominated by $c$.

\section{Analysis of DisDCA for Orthogonal Data}\label{sec:3}
In this section, we present our first theoretical result regarding the convergence rate of the practical DisDCA for orthogonal data on different machines. Actually, in this case the practical DisDCA can be modified slightly to obtain better convergence.  We can set  $scale=1$ in {\bf Inc-Dual}, i.e., updating $\alpha$ by
\begin{align}\label{eqn:dual-o}
 \Delta\alpha_{k,i}&= \max_{\Delta\alpha}-\phi_{k,i}^*(-(\alpha_{k,i}^{t-1} + \Delta\alpha)) - \Delta\alpha x_{k,i}^{\top}{u_k^{t, j-1}}\nonumber\\
 & \hspace*{0.4in}- \frac{1}{2\lambda n}(\Delta\alpha)^2\|x_{k,i}\|_2^2
 \end{align}
Accordingly, we need to change the updates of the primal variables $u^{t,j}_k$ to 
\begin{align}\label{eqn:u}
u^{t, j}_k &= u^{t, j-1}_k + \frac{1}{\lambda n}\Delta\alpha_{k,i}x_{k,i}
\end{align}
As we can see the $scale$ factor is reduced to $1$ compared to $K$ in the original DisDCA algorithm of the practical variant. This is due to the exploitation of the orthogonality of data on different machines. The theorem below present the convergence rate of the practical DisDCA in this case. 

\begin{thm}\label{thm:2}
Assume all data points are in the unit ball, i.e.,  $\|x\|_2\leq 1$, and  the data $X_k$ on different machines are orthogonal to each other, i.e., $X_k^{\top}X_l=0, \forall k\neq l$. 
For a $L$-smooth loss function $\phi_i$ and a $\lambda$-strongly convex function $g(w)$, let $w^T, \alpha^T$ be the primal and dual solution obtained by the practical DisDCA algorithm with slight modifications $scale=1$ and~(\ref{eqn:u}), then 
\begin{align*}
\E\left[D(\alpha^*) - D(\alpha^T)\right] \leq \left(1 - \frac{K}{c +n}\right)^{mT}\epsilon_0
\end{align*}
and 
\begin{align*}
\E\left[P(w^T) - D(\alpha^T)\right] \leq \frac{c + n}{K}\left(1 - \frac{K}{c + n}\right)^{mT}\epsilon_0
\end{align*}

\end{thm}
{\bf Remark:} Theorem~\ref{thm:2} well justifies  the benefit of the practical DisDCA compared with the naive DisDCA. In particular, increasing $m$ can always speed-up the convergence rate by an exponential rate
 . Also, there is a linear speed-up by increasing the number of machines.  In addition, the convergence rate in Theorem~\ref{thm:2} for the practical DisDCA is better than that in Theorem~\ref{thm:1} for the naive DisDCA. 

To understand the theoretical result in Thereom~\ref{thm:2}, we note that when the data on different machines are orthogonal, the Gram matrix $G= [x_i^\top x_j]_{n\times n}$ can be shuffled to align the distribution of data on $K$ machines, and becomes a block diagonal matrix 
\[
G = \left[\begin{array}{cccc}G_1&0&\cdots&0\\0&G_2&\cdots&0\\\cdot&\cdot&\cdot&\cdot\\ 0&\cdots&0&G_k\end{array}\right]
\]
and the dual problem in~(\ref{eqn:dual}) with a square norm regularizer can be split into $K$ independent sub-problems:
\begin{align*}
&\max_{\alpha\in\R^n} \frac{1}{n} \sum_{i=1}^n-\phi_i^*(-\alpha_i)  - \frac{1}{2n^2\lambda}\alpha^{\top} G \alpha\\
& = \max_{\alpha\in\R^n}\frac{1}{n} \sum_{k=1}^K\sum_{i=1}^{n_k}-\phi_i^*(-\alpha_i)  - \frac{1}{2n^2\lambda}\sum_{k=1}^K\alpha_k^{\top}G_k\alpha_k
\end{align*}
As we see shortly, an algorithmic consequence due to the orthogonality is that the communication between $K$ machines becomes optional (c.f. equation in~(\ref{eqn:oth})), as a result  we solve $K$ independent sub-problems separately. Therefore, increasing $m$ the number of dual variables to be updated on each machine would yield an exponential speed-up and increasing $K$ the number of machines would yield a linear speed-up. 

Although we note that it is usually  rare to have the property of orthogonality in reality~\footnote{The orthogonality may occur when data on different machines have non-overlapping features.}, however, the discussion here can be  related to a particular type of distributed optimization algorithms, i.e., one-communication  distributed optimization. In particular, one can regard the data on different machines are orthogonal and solve $K$ sub-problems separately and finally perform an average of models from all machines. It was advocated by practitioners and has been analyzed in~\citep{citeulike:11324985, citeulike:6779152,DBLP:conf/nips/ZinkevichWSL10} for least square regression, conditional maximum entropy models and stochastic gradient descent. However, the final averaged solution may not be (in most cases) exactly the optimal solution to the original problem and our analysis reveals that the orthogonality plays an important role in the accuracy of the final averaged solution.  Moreover, it suggests the closer to orthogonality of data the closer to the optimal solution of the averaged solution. 

\begin{figure}[t]
\center
\subfigure[residual of model]{\includegraphics[scale=0.2]{{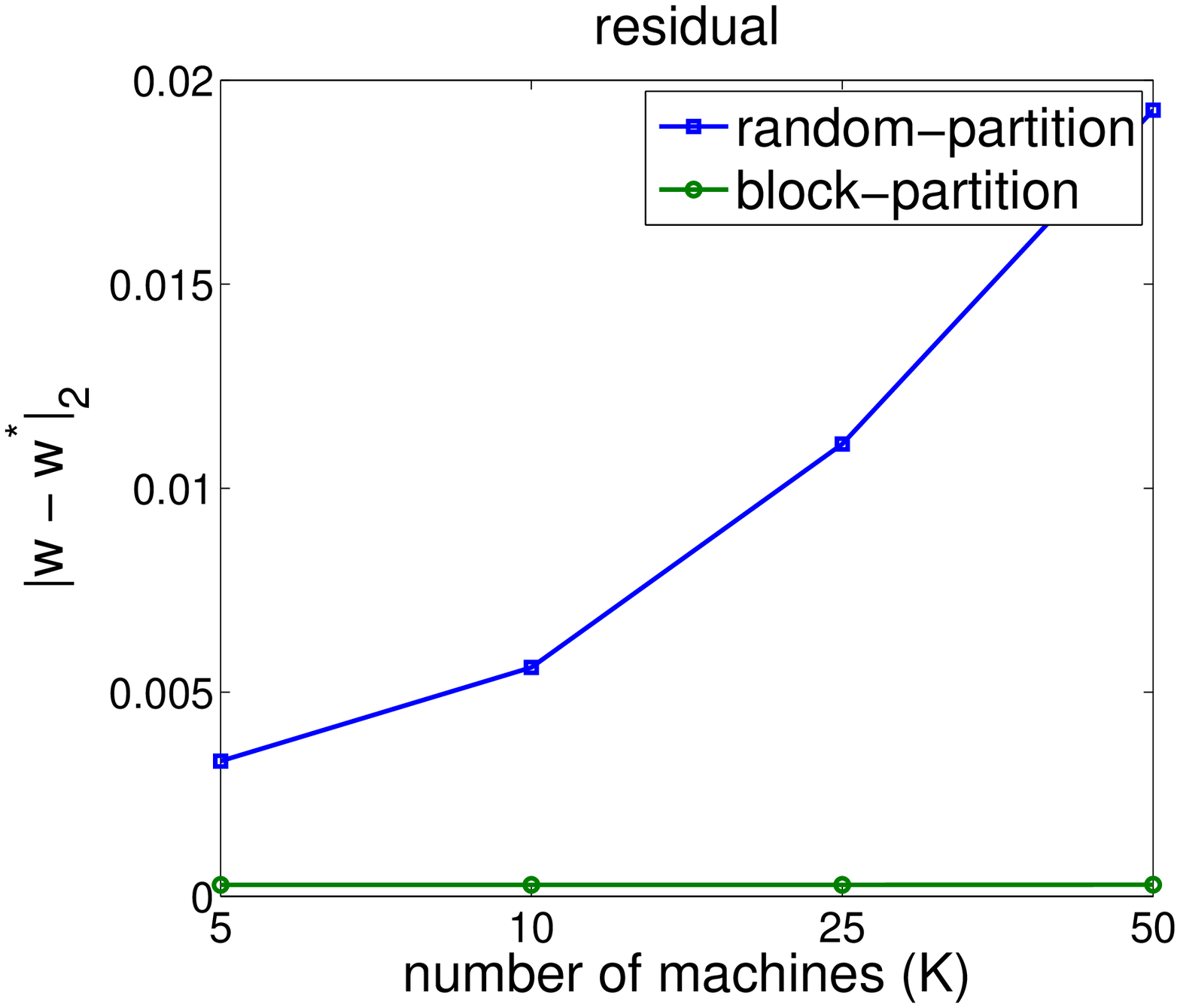}}}
\subfigure[duality gap]{\includegraphics[scale=0.2]{{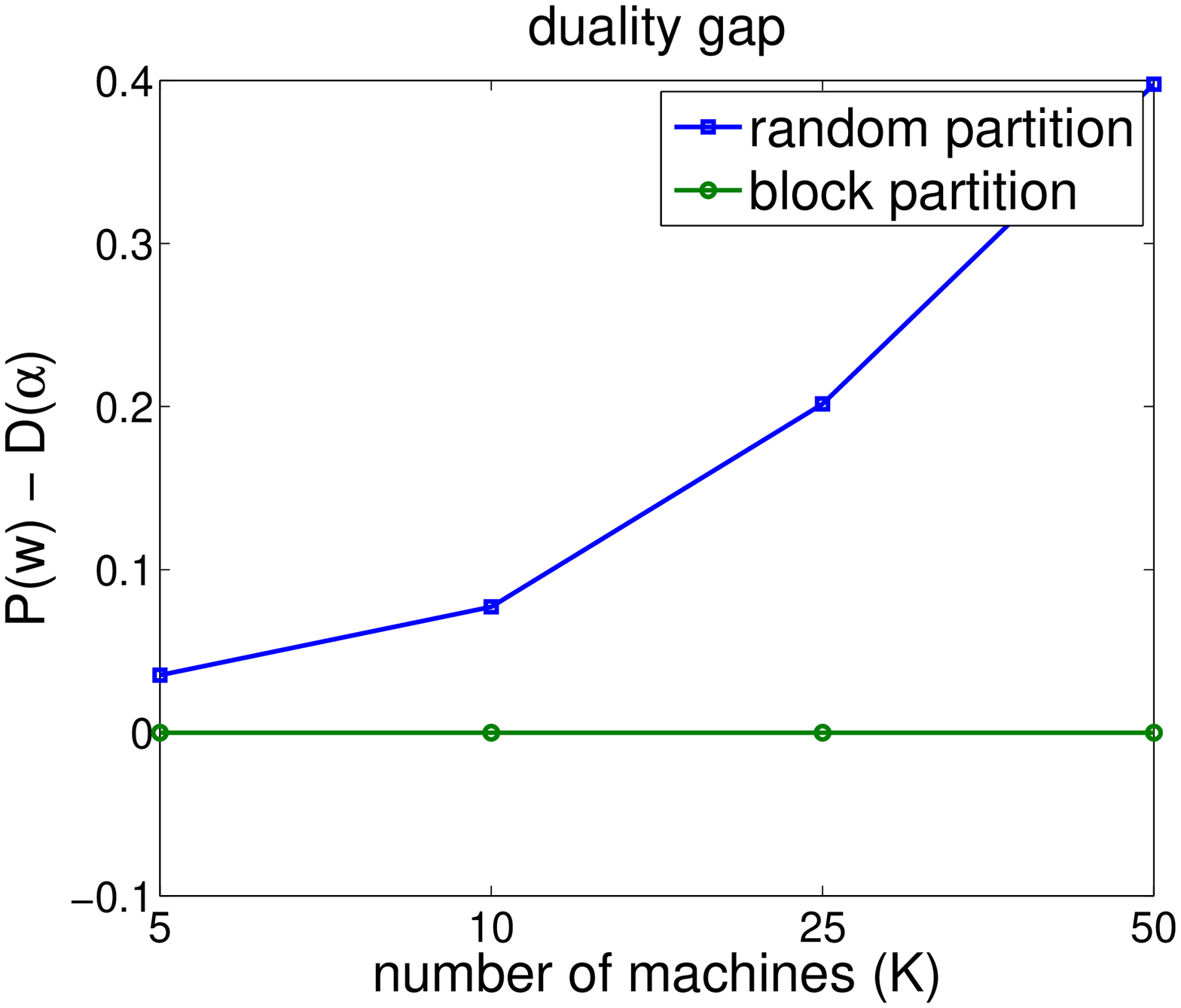}}}
\caption{Random partition vs Block Partition for  one-communication DisDCA. (a) shows the curves of the distance from the final averaged solution to the optimal solution; (b) shows the curves of the duality gap of the final averaged solution for the original problem. The value of $\lambda$ is set to $10^{-5}$. For each $K$, we run dual updates until the duality gaps for the sub-problems are within $10^{-6}$. The optimal solution $w^*$ is obtained by running the practical DisDCA until the dually gap is within $10^{-8}$.}\label{fig:partition}
\end{figure}

To verify our claim, we present an experimental result in Figure~\ref{fig:partition}, which compares the one-communication DisDCA with different data partitions on a synthetic data for regression. The data is generated similarly as in~\citep{citeulike:11324985} with extra concern.  In particular, we generate data of $d=250$ dimension and split the $250$ features into $50$ non-overlapping groups. For each group, we generate $5$ features following $\mathcal N(0, 1)$ for $5000$ data points. The response variable is generated by $y= u^{\top}x + \sum_{j=1}^d(x_j/2)^3$, where $u$ is a constant vector. As a result, there are a total of $250000$ data points. We distribute the data over $K=5, 10, 25, 50$ machines. For each $K$, we generate two data partitions, one following the block partition and another one following a random partition. For the block partition, we run the updates in~(\ref{eqn:dual-o}), for the random partition we run the updates in~(\ref{eqn:dual-p}). We note that for the random partition, the update in~(\ref{eqn:dual-o}) would fail. For each sub-problem, we run the dual updates until the duality gaps of the sub-problems are within $10^{-6}$. 
The results verify that the one-communication DisDCA for orthogonal data partition achieves better optimality. 
\begin{figure}[t]
\center
\subfigure[comparison of two variants]{\includegraphics[scale=0.2]{{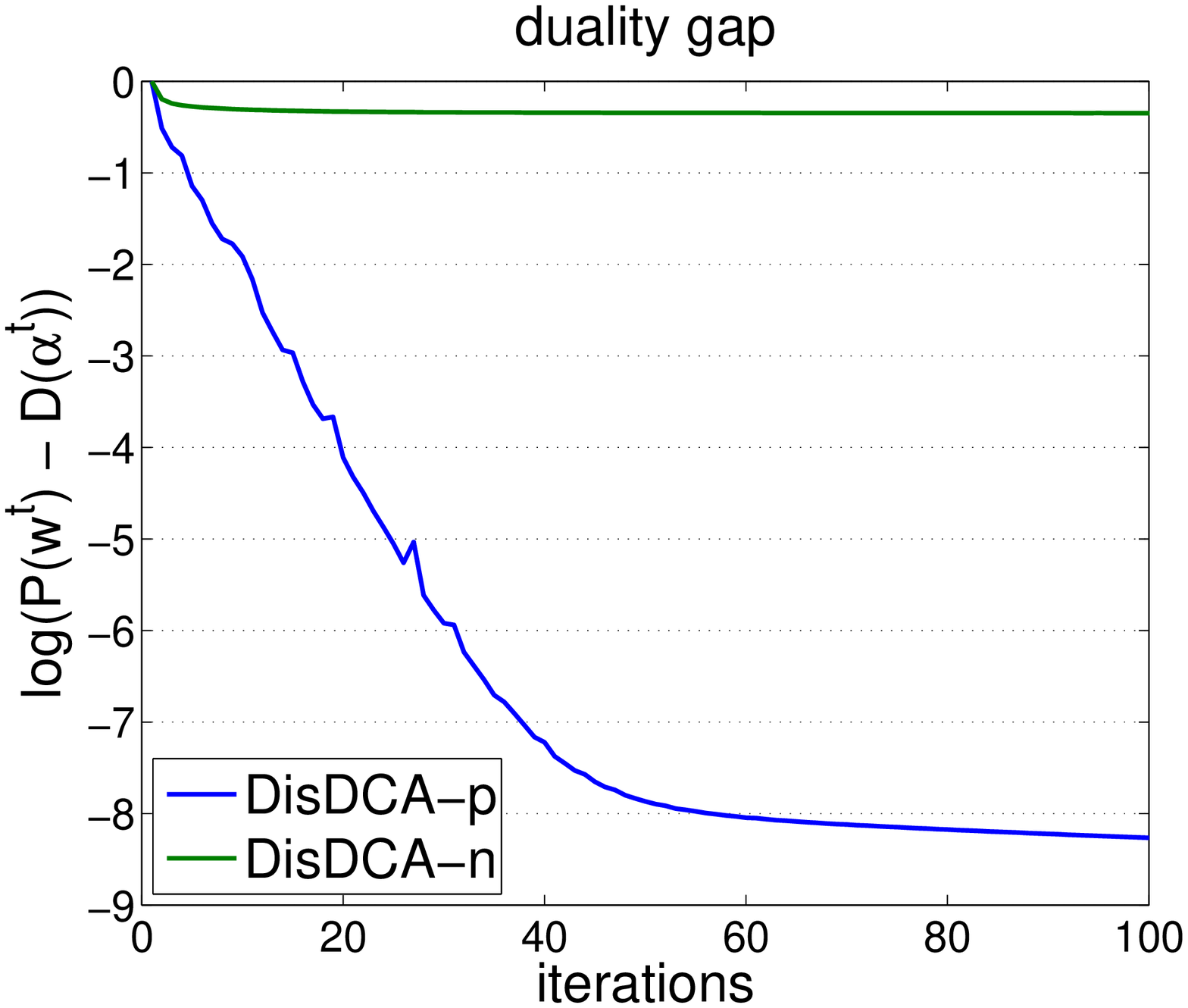}}}
\subfigure[running time and error ]{\label{fig:rt}\includegraphics[scale=0.2]{{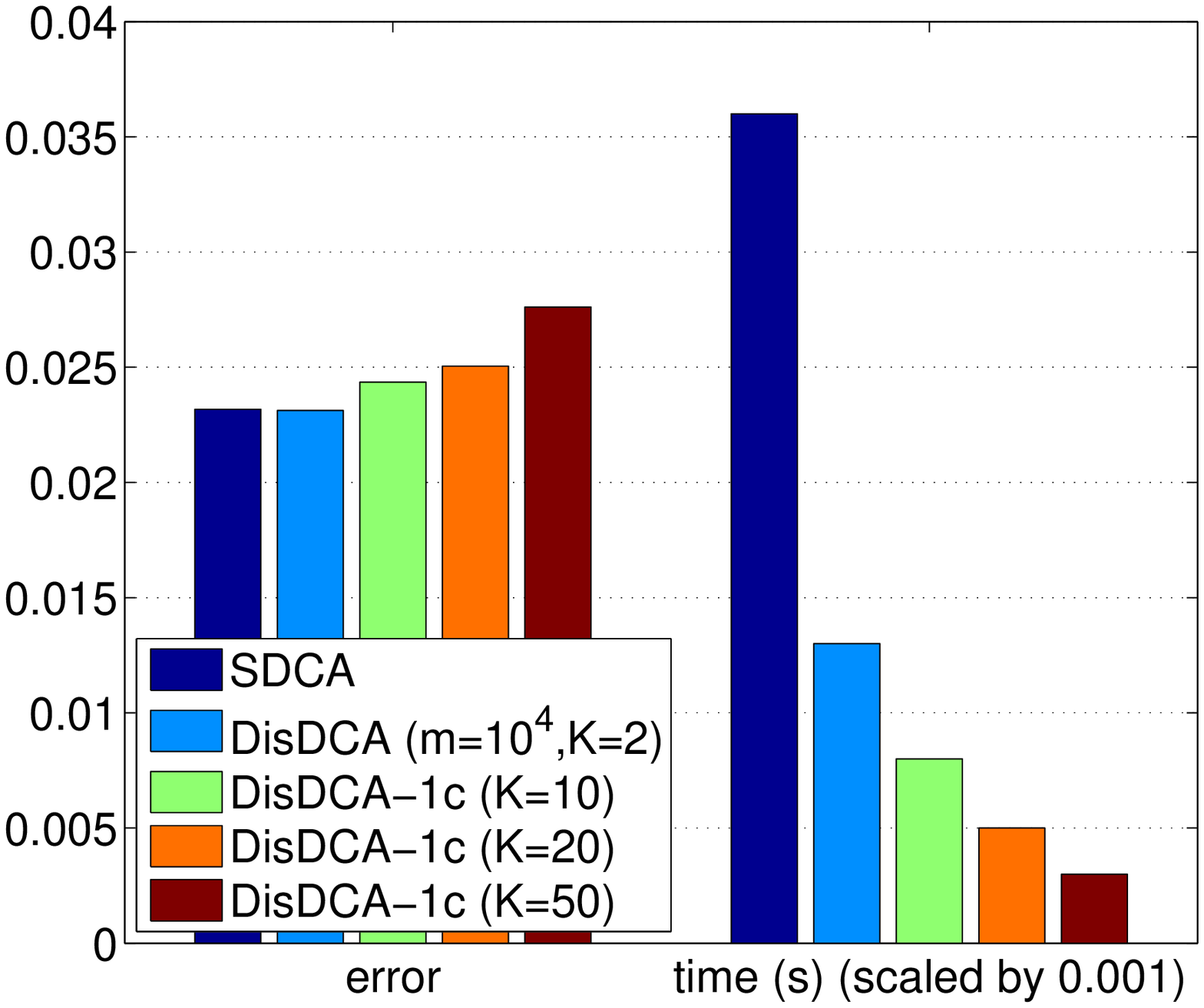}}}
\caption{(a): comparison of the practical DisDCA (DisDCA-p) with the naive DisDCA (DisDCA-n) on covtype data set with $m=1000, K=10$; (b) comparison of running time and error rate for SDCA, DisDCA and one-communication DisDCA (DisDCA-1c) on rcv1-binary data set. Running time excludes the time of loading data. The loss function is the squared hinge loss in both experiments and the value of $\lambda$  is set to $10^{-6},10^{-5}$, respectively.}\label{fig:1}
\end{figure}

Finally, we note that although the one-communication DisDCA optimization with a random partition may not find the optimal solution, however, it may give sufficiently accurate predictions. For example, in Figure~\ref{fig:rt} we show the error of  classification~\footnote{We use the provided testing data ($n=677399$, $d=47236$) as training data and evaluate the model on the provided training data.} on rcv1 binary data set~\footnote{\url{http://www.csie.ntu.edu.tw/~cjlin/libsvmtools/datasets/}}. The results show that with $K=20$ machines, the one-communication DisDCA gains $8$-times speed-up in running time while only losses $0.2\%$ in accuracy. Similar results have been reported in previous works~\citep{citeulike:11324985, citeulike:6779152}, where they have established the statistical convergence of the final averaged solution as well.  Our analysis in next section reveals the optimization convergence of the one-communication DisDCA. 




\subsection{Proof} 
To ease the proof, 
we introduce some notations to simplify our analysis.  
Let $\alpha^{t, j}$ denote all  dual variables at the $t$-th iteration after updating the $j$-th variable, $\alpha^t = \alpha^{t, 0}$ and $\alpha^{t+1} = \alpha^{t, m}$.   Let $w^t_k$ denote the local primal solution spans over the data points in its corresponding machine, i.e., 
\begin{align*}
w_k^{t} =  \frac{1}{\lambda n}\sum_{i=1}^{n_k}\alpha_{k,i}^t x_{k,i}, \quad w^t  = \sum_{k=1}^Kw^t_k
\end{align*}
Note that compared to Algorithm~\ref{alg:disdca} we slightly change the definition of  $w_k^t$. 

Similarly, let $w^{t,j}_k$ and $w^{t,j}$ denote the local and global primal solution constructed from the updated dual variables $\alpha^{t,j}$ after updating the $j$-th dual variable at the $t$-th iteration, i.e.,  
\begin{align*}
&w_k^{t, j} = w_k^{t, j-1} + \frac{1}{\lambda n}\Delta\alpha_{k, i_j}^{t, j}x_{k, i_j},\quad w_k^{t,0}= w_k^{t-1},\\
&w^{t,j} = \sum_{k=1}^Kw^{t,j}_k
\end{align*}
Thus, we can establish the relationship between $w^{t,j}_k$ and $u^k_{t,j}$ as follows: 
\begin{align*}
u_k^{t, j} = u_k^{t,j-1} + \frac{1}{\lambda n}\Delta\alpha^{t, j}_{k, i_j}x_{k, i_j} = w_k^{t, j} + \bar w_k^{t-1}
\end{align*}
where $\bar w_k^{t-1} = w^{t-1} - w_k^{t-1}$. Essentially,  $w^{t,j}$ denotes the summed solution of all local $w^{t,j}_{k}$, which is not computed in the algorithm. $u^{t,j}_k$ denotes the local variable maintained at each machine, whose starting point at the beginning of the inner iteration is the global $w^{t-1}$. Due to that the data on different machines are orthogonal, it is not difficult to verify that
\begin{align}
 &x_{k,i}^{\top}w^{t, j-1}=x_{k,i}^{\top}w_k^{t, j-1}=x_{k,i}^{\top}u_k^{t,j-1}\label{eqn:oth}\\
 &\left\|\sum_{k=1}^K w_k^t\right\|_2^2 = \sum_{k=1}^K\|w_k^t\|_2^2\label{eqn:decomp}.
 \end{align}
An important consequence springing from the data orthogonality is that we can update the dual variables separately without any loss in decomposing the quadratic term in the dual objective~(\ref{eqn:dual}), i.e., 
\begin{align*}
&\frac{\lambda n}{2}\left\|w^{t,j-1} + \frac{1}{\lambda n}\sum_{k}\Delta\alpha_{k,i_j}x_{k, i_j}\right\|^2\\
&=\frac{\lambda n}{2}\|w^{t,j-1}\|_2^2 + \sum_{k=1}^K\frac{1}{\lambda n}(\Delta\alpha_{k,i_j})^2\|x_{k,i_j}\|_2^2+ \sum_{k=1}^K\Delta\alpha_{k,i_j}x_{k,i_j}\cdot u_k^{t, j-1}
\end{align*}
The last equality is due to~(\ref{eqn:oth}). This is exactly where the update in~(\ref{eqn:dual-o}) comes from. 


Below, we divide the proof into three parts:  (1) bounding the increase of the dual objective for one inner iteration; (2) establishing the relationship between the increased dual objective and the duality gap; and (3) finally proving the convergence rate of the dual objective and the duality gap. For each part, we present a Lemma with the proof deferred to appendix. 

We start by bounding the increase of the dual objective for one inner iteration.  It is easy to verify that $w^{t,j} = \frac{1}{\lambda n}\sum_{k,i}\alpha^{t,j}_{k,i}x_{k,i}$. By the definition of $D(\alpha)$, we have 
\begin{align*}
&n[D(\alpha^{t, j}) - D(\alpha^{t, j-1})] \\
&= \left[\sum_{k}\sum_{i}-\phi^*_{k,i}(-\alpha_{k,i}^{t,j}) - \frac{\lambda n}{2}\|w^{t, j}\|_2^2\right] -  \left[\sum_{k}\sum_{i}-\phi^*_{k,i}(-\alpha_{k,i}^{t,j-1}) - \frac{\lambda n}{2}\|w^{t, j-1}\|_2^2\right]\\
&= \sum_k\underbrace{\left[ - \phi^*_{k,i_j}(-\alpha^{t, j}_{k,i_j}) - \frac{\lambda n}{2}\|w_k^{t,j}\|_2^2\right]}\limits_{A_k}
-\sum_k\underbrace{\left[ - \phi^*_{k,i_j}(-\alpha^{t, j-1}_{k,i_j}) - \frac{\lambda n}{2}\|w_k^{t,j-1}\|_2^2\right]}\limits_{B_k}
\end{align*}
The last equality is because we only update the dual variable of the data point with index $i_j$, and the equality in~(\ref{eqn:decomp}). 
We proceed to bound $A_k$ by writing $\alpha_{k,i_j}^{t,j} = \alpha_{k,i_j}^{t,j} + \Delta\alpha$ and $w_k^{t,j}= w_k^{t,j-1} + 1/(\lambda n)\Delta\alpha_{k,i_j} x_{k,i_j}$, and bounding $\phi^*(\alpha)$ using its strong convexity. The result is summarized in the following Lemma. 
\begin{lemma} Let $\omega_{k,i_j}^{t, j-1}= -\nabla \phi(x_{k,i_j}\cdot w^{t, j-1}_k)$. Then for any $s\in[0,1]$, we have
\begin{align*}
&A_k- B_k\geq \frac{s}{2}\left(\frac{1-s}{L} - \frac{s\|x_{k,i_j}\|_2^2}{\lambda n}\right)\|\omega_{k,j_j}^{t, j-1}-\alpha^{t,j-1}_{k, i_j}\|_2^2\\
&+ s\left[\phi_{k,i_j}(x_{k,i_j}\cdot w_k^{t,j-1})+\phi^*_{k,i_j}(-\alpha^{t,j-1}_{k,i_j})+\alpha_{k, i_j}^{t, j-1}x_{k,i_j}^{\top}w^{t, j-1}_{k}\right].
\end{align*}
\end{lemma}
Since we assume $\|x\|_2\leq 1$, we can set  $\displaystyle s = \frac{n}{ c + n}$ and we have
\begin{align*}
&A_k - B_k \geq s\left[\phi_{k,i_j}(x_{k,i_j}\cdot w_k^{t,j-1})+\phi^*_{k,i_j}(-\alpha^{t,j-1}_{k,i_j})+\alpha_{k, i_j}^{t, j-1}x_{k,i_j}^{\top}w^{t, j-1}_{k}\right].
\end{align*}
Taking summation over $k=1,\ldots, K$ on both sides and taking expectation over the randomness, we can prove the following Lemma, which establishes the relationship between the increased dual objective and the duality gap. 
\begin{lemma} \label{lem:2} Let $s = n/(c+n)$. Then
\begin{align*}
&\hspace*{-0.5in}\E\left[D(\alpha^{t,j}) - D(\alpha^{t,j-1})\right]\geq \frac{sK}{n}\left[P(w^{t,j-1}) - D(\alpha^{t,j-1})\right].
\end{align*}
\end{lemma}
Given Lemma~\ref{lem:2} and $D(\alpha^*)\leq P(w^{t, j-1})$, we have
  \begin{align*}
 &\hspace*{-0.2in}\E\left[D(\alpha_*) - D(\alpha^{t})\right]\leq \left(1 - \frac{sK}{n}\right)^m \E\left[D(\alpha_*) - D(\alpha^{t-1})\right]
 \leq  \left(1 - \frac{sK}{n}\right)^{mt}\epsilon_0
 &
 \end{align*}
 Therefore
  \begin{align*}
 \E\left[D(\alpha_*) - D(\alpha^{T})\right]&\leq  \left(1 - \frac{sK}{n}\right)^{mT}\epsilon_0= \left(1 - \frac{K}{c + n}\right)^{mT}\epsilon_0
 \end{align*}
 and
 \begin{align*}
 \E\left[P(w^{T}) - D(\alpha^{T})\right]&\leq \frac{n}{sK}\E\left[D(\alpha_*) - D(\alpha^{T})\right]=\frac{c+n}{K} \left(1 - \frac{K}{c + n}\right)^{mT}\epsilon_0
 \end{align*}
 
 \section{Analysis of the Practical DisDCA in General Case}
 The challenge for the analysis of  the general case comes from that we cannot decompose the square norm of $w^{t,j}$ as the sum of square norm of each $w^{t,j}_k$ without any loss. Our  strategy for analysis is to derive a similar inequality as in Lemma~\ref{lem:2}. However, as we will see shortly, there is an additional term that accounts for the difference between using the global primal solution $w^{t, j-1}$ and the local primal solution $u^{t, j-1}_k$.  According to the updates, let us define:
 \begin{align*}
&w_k^{t} =  \frac{K}{\lambda n}\sum_{i=1}^{n_k}\alpha_{k,i}^t x_{k,i}, \quad w^t  = \frac{1}{K}\sum_{k=1}^Kw^t_k, \\
&w_k^{t, j} = w_k^{t, j-1} + \frac{K}{\lambda n}\Delta\alpha_{t, i_j}^{t, j}x_{k, i_j},\quad  w^{t,j} = \frac{1}{K}\sum_{k=1}^Kw^{t,j}_k,\\
&w_k^{t, 0}= w_k^{t-1}, \quad w^{t+1} = w^{t, m},\quad u_k^{t,0} = w^{t-1}\\
&u_k^{t, j} = u_k^{t,j-1} + \frac{K}{\lambda n}\Delta\alpha^{t, j}_{k, i_j}x_{k, i_j} = w_k^{t, j} + \bar w_k^{t-1}
\end{align*}
where $\bar w_k^{t-1} = w^{t-1} - w_k^{t-1}$. 
The following lemma establishes a similar result as in Lemma~\ref{lem:2}. 
 \begin{lemma}\label{lem:3}Let $\omega^{t,j-1}_{k, i} = -\nabla\phi(x_{k,i}\cdot w^{t, j-1})$.  For any $s\in[0,1]$, we have
\begin{align*}
&\E[D(\alpha^{t, j}) - D(\alpha^{t, j-1})]\geq \frac{sK}{n}\E[P(w^{t,j-1}) - D(\alpha^{t, j-1})]\nonumber\\
&+ s\E\sum_{k=1}^K\left(\frac{1-s}{2L}- \frac{sK}{2\lambda n}\|x_{k,i_j}\|_2^2\right)\|\omega^{t,j-1}_{k, i_j}-\alpha^{t,j-1}_{k, i_j}\|_2^2 - \E\left[R^{t,j}\right],
\end{align*}
where
\begin{align*}
&R^{t,j}=\frac{1}{n} \sum_{k=1}^K(s(\omega^{t,j-1}_{k, i_j}-\alpha^{t,j-1}_{k, i_j})- \Delta\alpha^{t, j}_{k, i_j} )x_{k,i_j}^{\top}(u_k^{t, j-1}  - w^{t,j-1})
\end{align*}
 \end{lemma}
 If we let $\displaystyle s =\frac{n}{cK + n}$, then the second term in the R.H.S of the  inequality in Lemma~\ref{lem:3} diminishes, and we get a similar result as in Lemma~\ref{lem:2}, except that there is an additional negative term $-R^{t,j}$. 
 
\begin{figure}[t]
\center
\subfigure[classification]{\includegraphics[scale=0.2]{{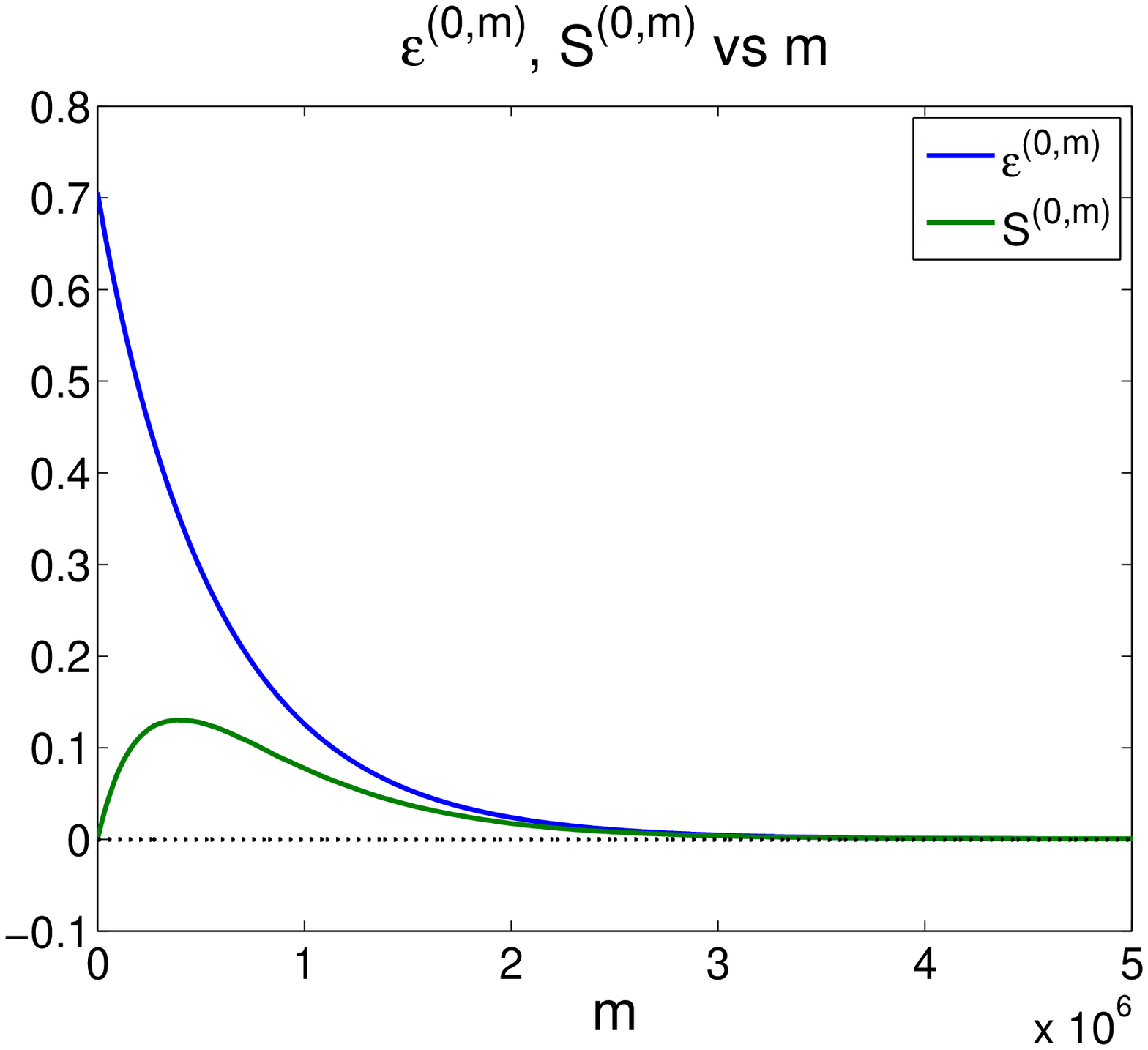}}}
\subfigure[regression]{\includegraphics[scale=0.2]{{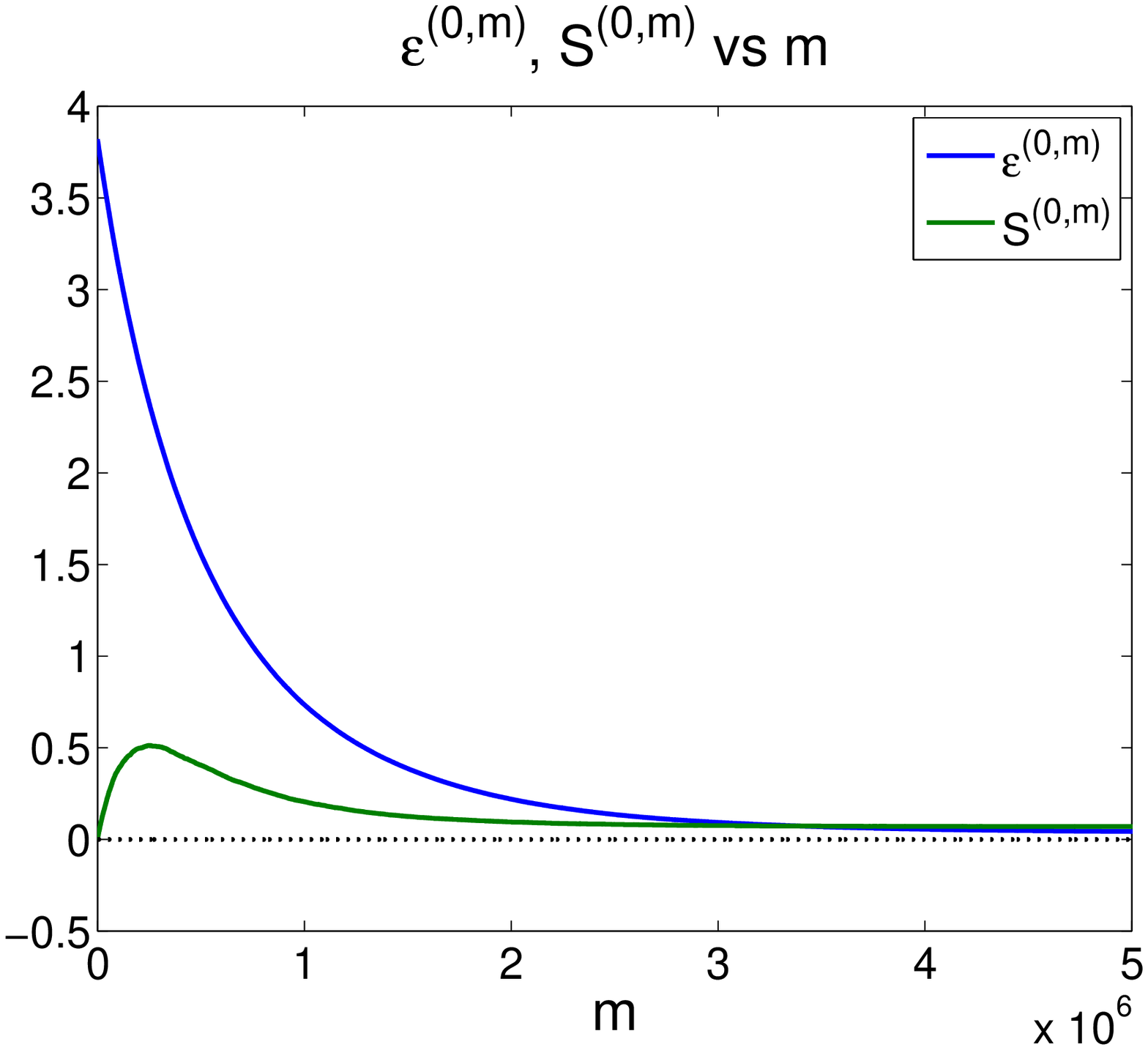}}}
\caption{Convergence of one-communication DisDCA. (a) shows the curves of $\epsilon^{(0, m)}$ and $S^{(0,m)}$ for classification using squared hinge loss on covtype data set; (b) shows the curves for regression using least square loss on the synthetic data as described in section~\ref{sec:3}. The convergent value for $\epsilon$ is slightly less than $S$. The value of $\lambda$ in both experiments are set to $10^{-5}$.}\label{fig:one-comm}
\end{figure}

 Ideally, if we use $w^{t,j-1}$ to replace $u_k^{t,j-1}$, then $R^{t,j}=0$ and we can prove a similar result. However, computing $w^{t,j-1}$ involves communication among all $K$ machines, which is not adopted in our algorithm. Essentially, $R^{t, j}$ can be regarded as a measure of deviation from the local primal solution $u_k^{t,j-1}$ to the global primal solution $w^{t, j-1}$. To see this, let use consider an example using squared hinge loss $\phi(x\cdot w) = (1-yx\cdot w)_+^2$, whose convex conjugate is $\phi^*(\alpha) = \alpha y + \frac{\alpha^2}{4}$.  It is safely to let $\Delta\alpha_{k,i_j}^{t,j}$ be the solution to the following problem, assuming $\|x\|_2\leq 1$,
\begin{align*}
\Delta\alpha^{t, j}_{k,i_j}  &= \max_{\Delta\alpha} - \phi^*_{k, i_j}(-\alpha^{t, j-1}_{k, i_j} - \Delta\alpha) - \Delta\alpha x_{k, i_j}^{\top}u_k^{t,j-1}- \frac{K}{2\lambda n}(\Delta\alpha)^2
\end{align*}
It is easy to check that $\Delta\alpha_{k, i_j}^{t,j}$ is given by 
\begin{align*}
\Delta\alpha^{t, j}_{k, i_j}  = \frac{\lambda n}{2K+\lambda n}\left[2(y_{k,i_j} - x_{k,i_j}^{\top}u^{t, j-1}_k) - \alpha^{t, j-1}_{k, i_j}\right]
\end{align*}
 Noting that $L=2, c=2/\lambda$, we have
 \begin{align*}
 s(\omega^{t,j-1}_{k,i_j} &- \alpha^{t, j-1}_{k,i_j}) = \frac{\lambda n}{2K + \lambda n}\left[2(y_{k, i_j} - x^{\top}_{k,i_j}w^{t, j-1}) -\alpha^{t, j-1}_{k, i_j}\right]
 \end{align*}
 Then we have
 \begin{align*}
 &R^{t, j} \leq \frac{1}{n}\sum_{k=1}^K\frac{\lambda n}{2K  + \lambda n}\|x_{k, i_j}\|_2^2\left\|u_k^{t, j-1} - w^{t, j-1}\right\|_2^2\\
 &\propto \sum_{k=1}^K\left\|\frac{K}{\lambda n}\sum_{\tau=1}^{j-1}\Delta\alpha^{t, \tau}_{k,i_\tau}x_{k,i_\tau} - \frac{1}{\lambda n}\sum_{k=1}^K\sum_{\tau=1}^{j-1}\Delta\alpha^{t, \tau}_{k,i_\tau}x_{k,i_\tau}\right\|_2^2
 \end{align*}
 When the dual variables $\alpha^{t,j}$ converge to the optimal solution $\alpha^*$, we have $\Delta\alpha^{t, j}\rightarrow 0$, therefore $R^{t, j}\rightarrow 0$. Following similar analysis, we have
 \begin{align*}
 \E[\epsilon^{(t,j)}]\leq \left(1 - \frac{sK}{n}\right)\E[\epsilon^{(t, j-1)}] + \E[R^{(t,j)}]
 \end{align*}
 Let $\mu = 1-(sK)/n$, and by induction we have
 \begin{align}\label{eqn:conv}
 \E[\epsilon^{(t, m)}] \leq \mu^m\E[\epsilon^{(t, 0)}] + \E\underbrace{\left[\mu^{m-1}R^{t,1} + \cdots + R^{t,m}\right]}\limits_{S^{(t,m)}}
 \end{align}
 Note that since we are mostly interested in the dependence  on $m$, we hide the dependence on $K$ in $S^{(t,m)}$. 
If we let $t=0, m\rightarrow \infty$, then the above bound indicates that the residual of the dual objective for the one-commutation DisDCA will geometrically converge to $S^{(0, \infty)}$. This coincides with the fact that one-communication DisDCA performs independent SDCA on individual machines .  In Figure~\ref{fig:one-comm}, we show that the curves of $\epsilon^{(0,m)}$ and $S^{(0, m)}$ for classification and regression.  The curves clearly illustrate that both  $\epsilon^{(0,m)}$ and $S^{(0, m)}$ will eventually converge to a value, where the convergent value of $\epsilon^{(0,m)}$ is slightly smaller than the convergent value of $S^{(0,m)}$. These results justify the 
the inequality in~(\ref{eqn:conv}) for one-communication DisDCA.  

 \begin{figure}[t]
\center
\hspace*{-0.2in}\subfigure[$m=10$]{\includegraphics[scale=0.18]{{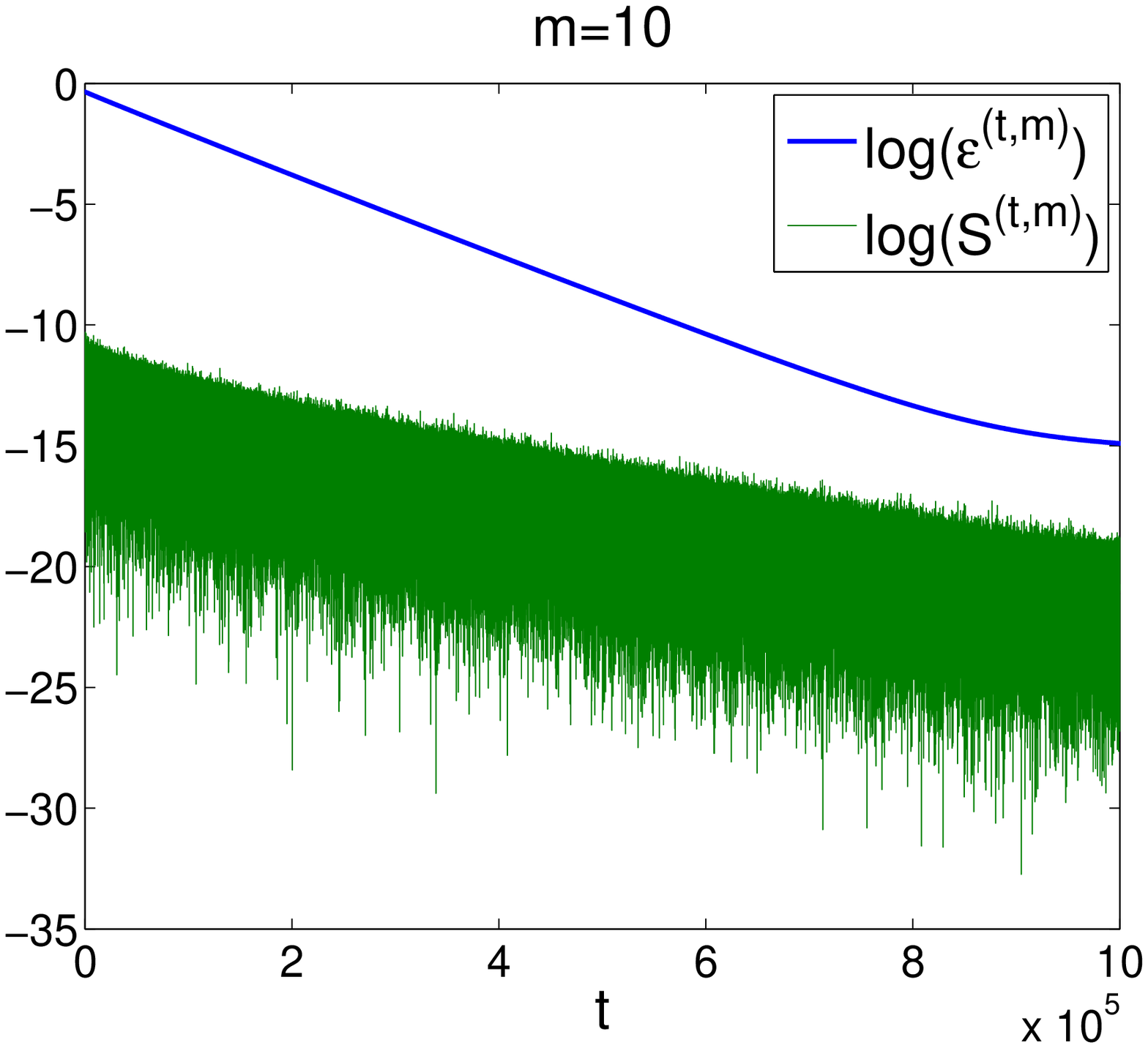}}}
\hspace*{-0.1in}\subfigure[$m=100$]{\includegraphics[scale=0.18]{{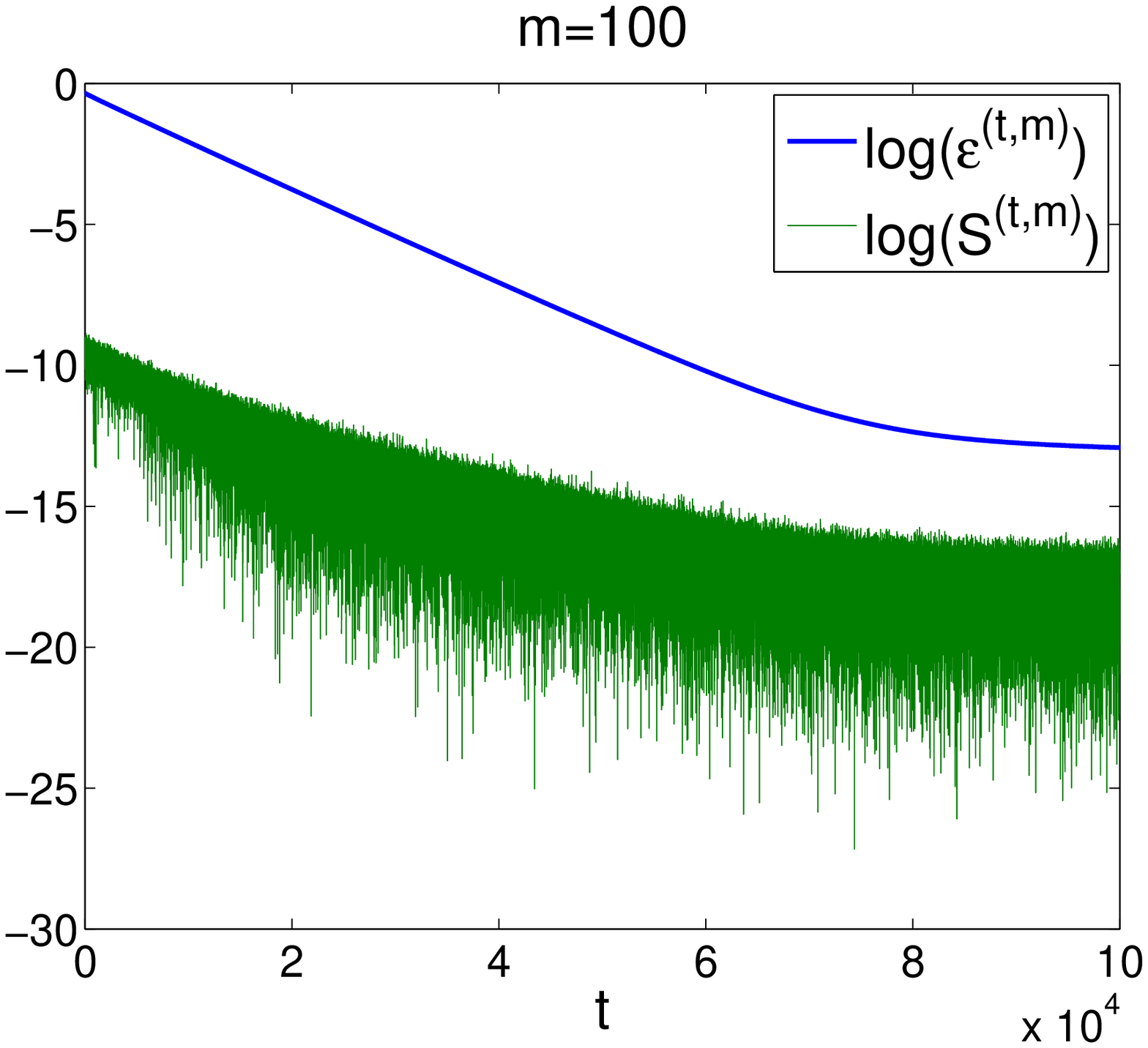}}}
\hspace*{-0.1in}\subfigure[$m=1000$]{\includegraphics[scale=0.18]{{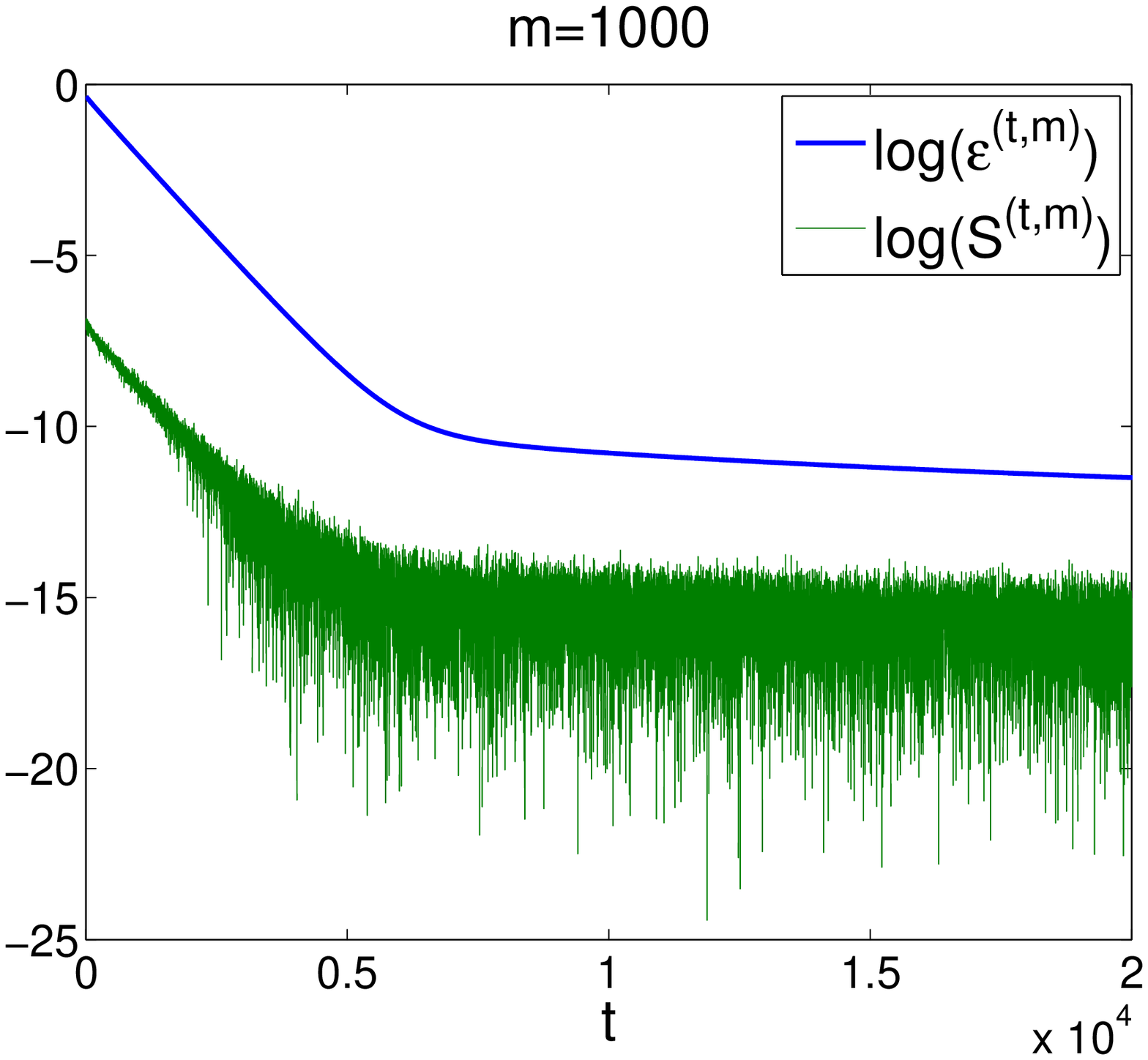}}}
\subfigure[varying $m$]{\includegraphics[scale=0.18]{{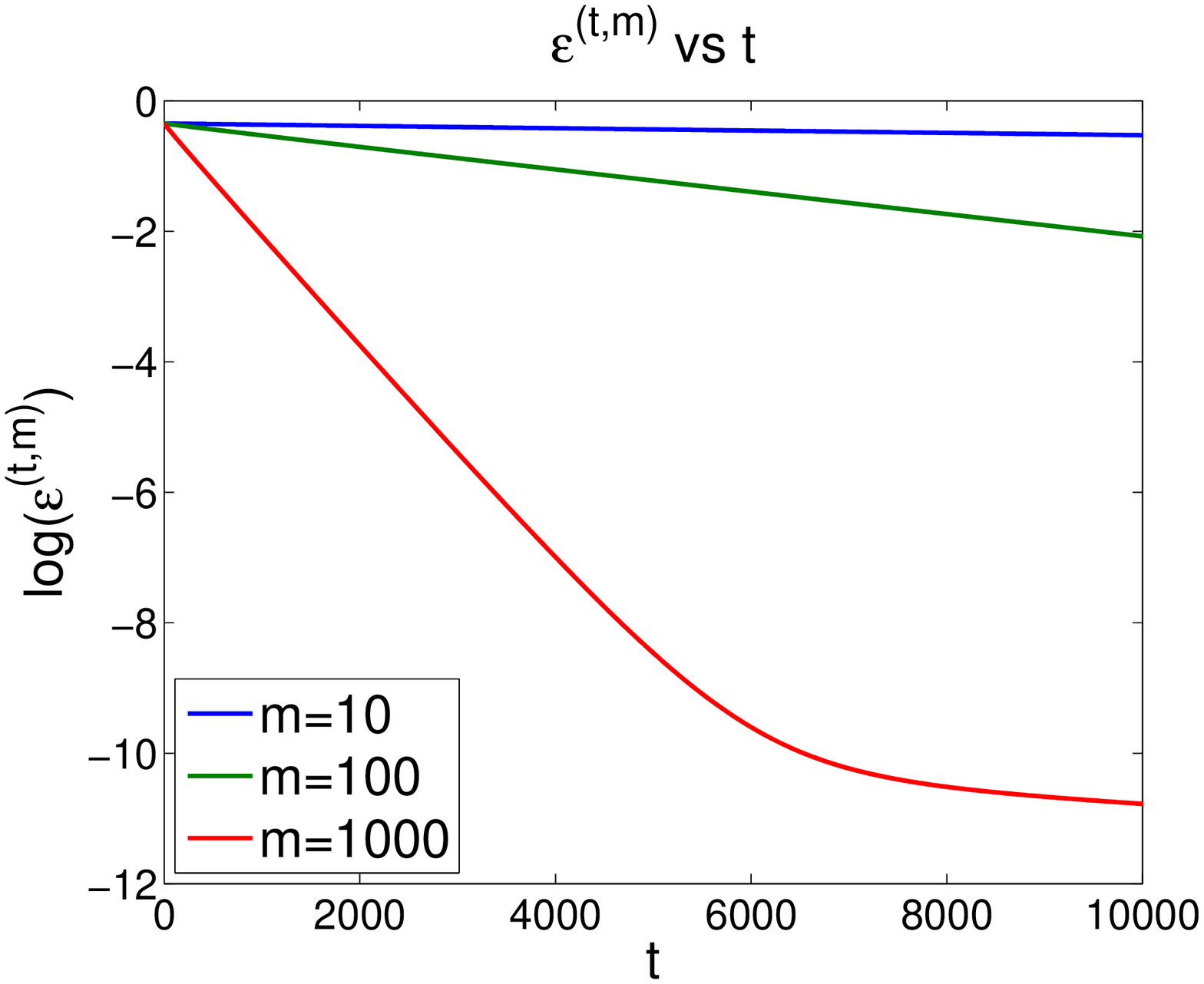}}}

\hspace*{-0.2in}\subfigure[$m=10$]{\includegraphics[scale=0.18]{{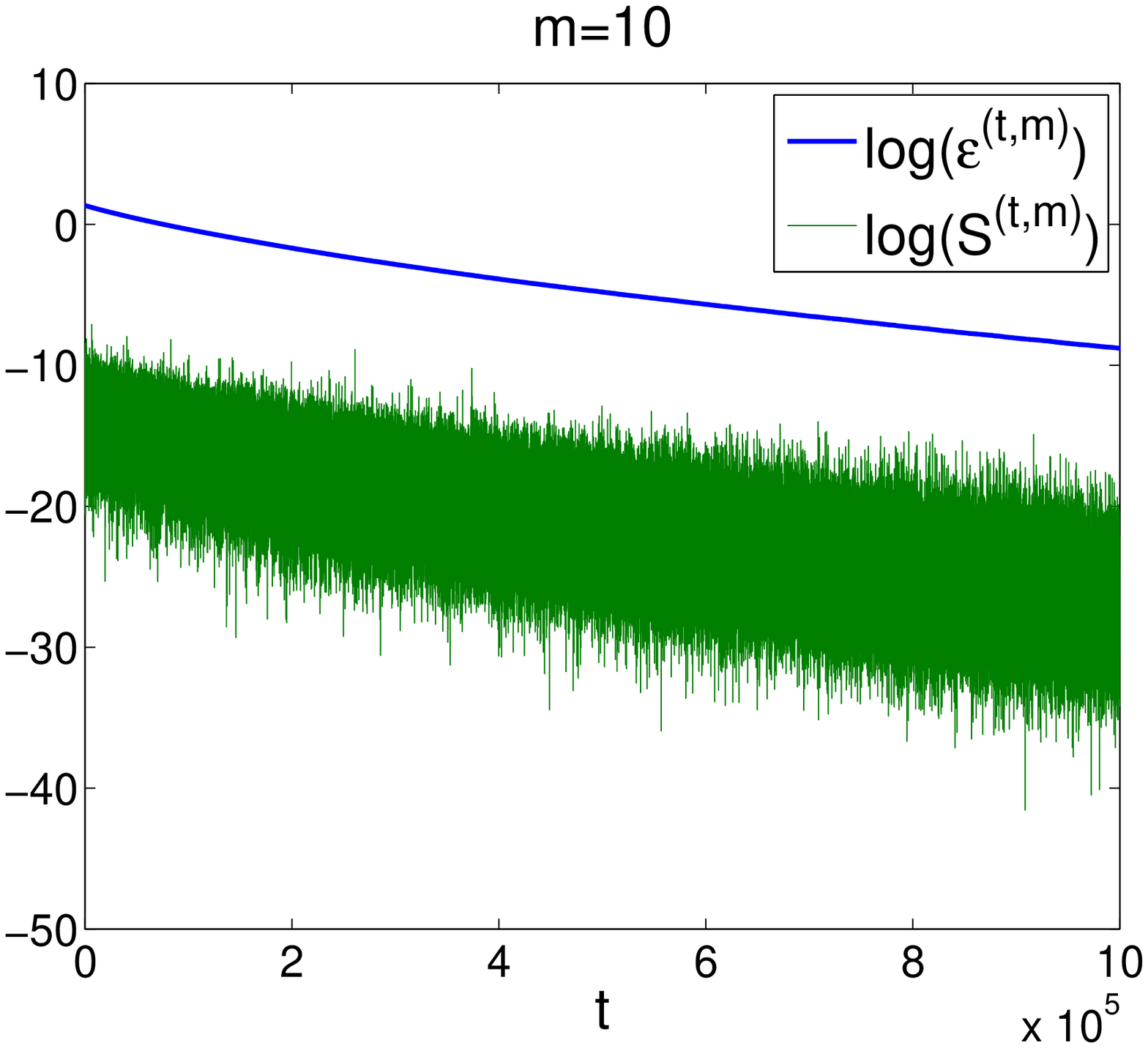}}}
\hspace*{-0.1in}\subfigure[$m=100$]{\includegraphics[scale=0.18]{{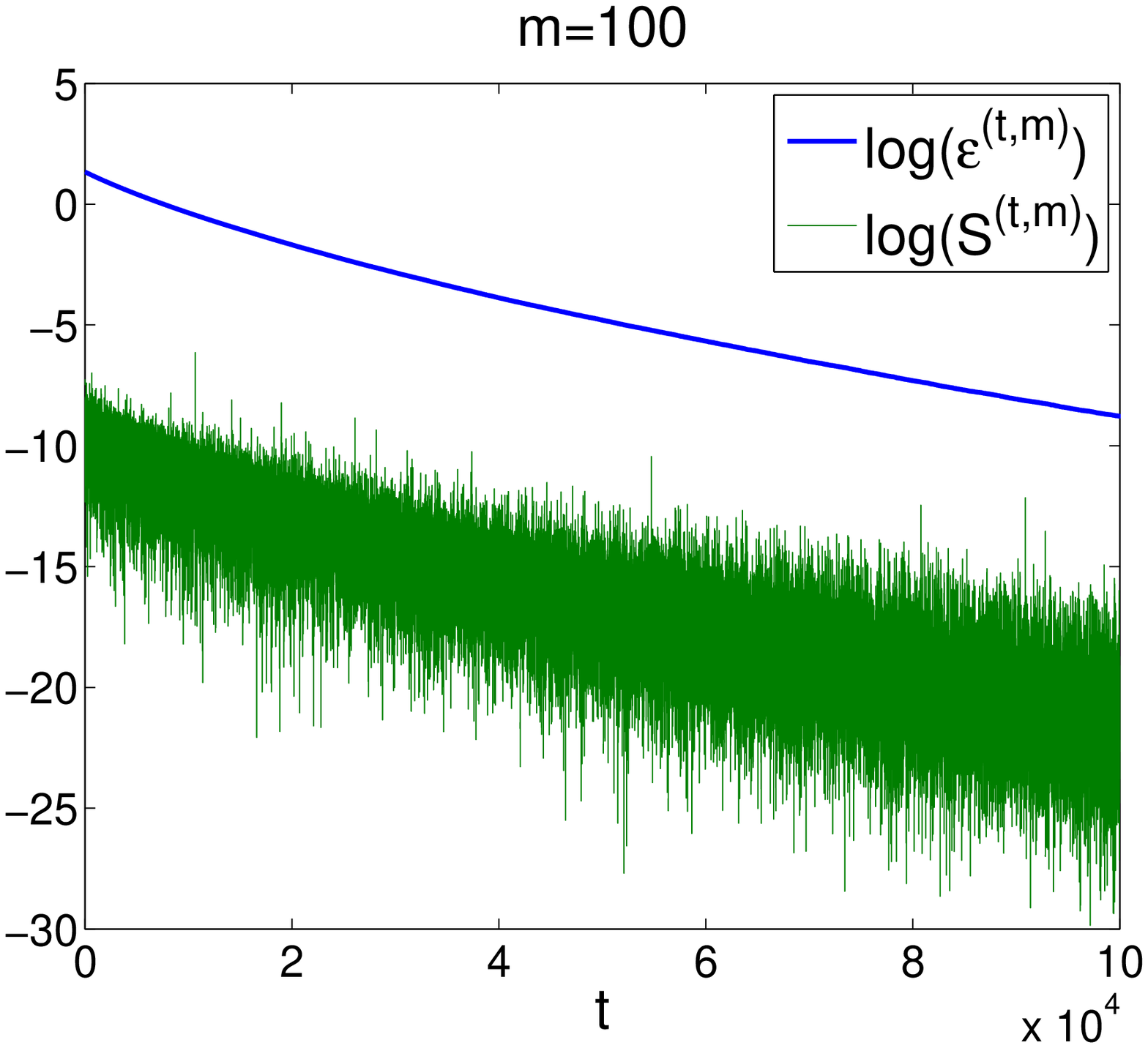}}}
\hspace*{-0.1in}\subfigure[$m=1000$]{\includegraphics[scale=0.18]{{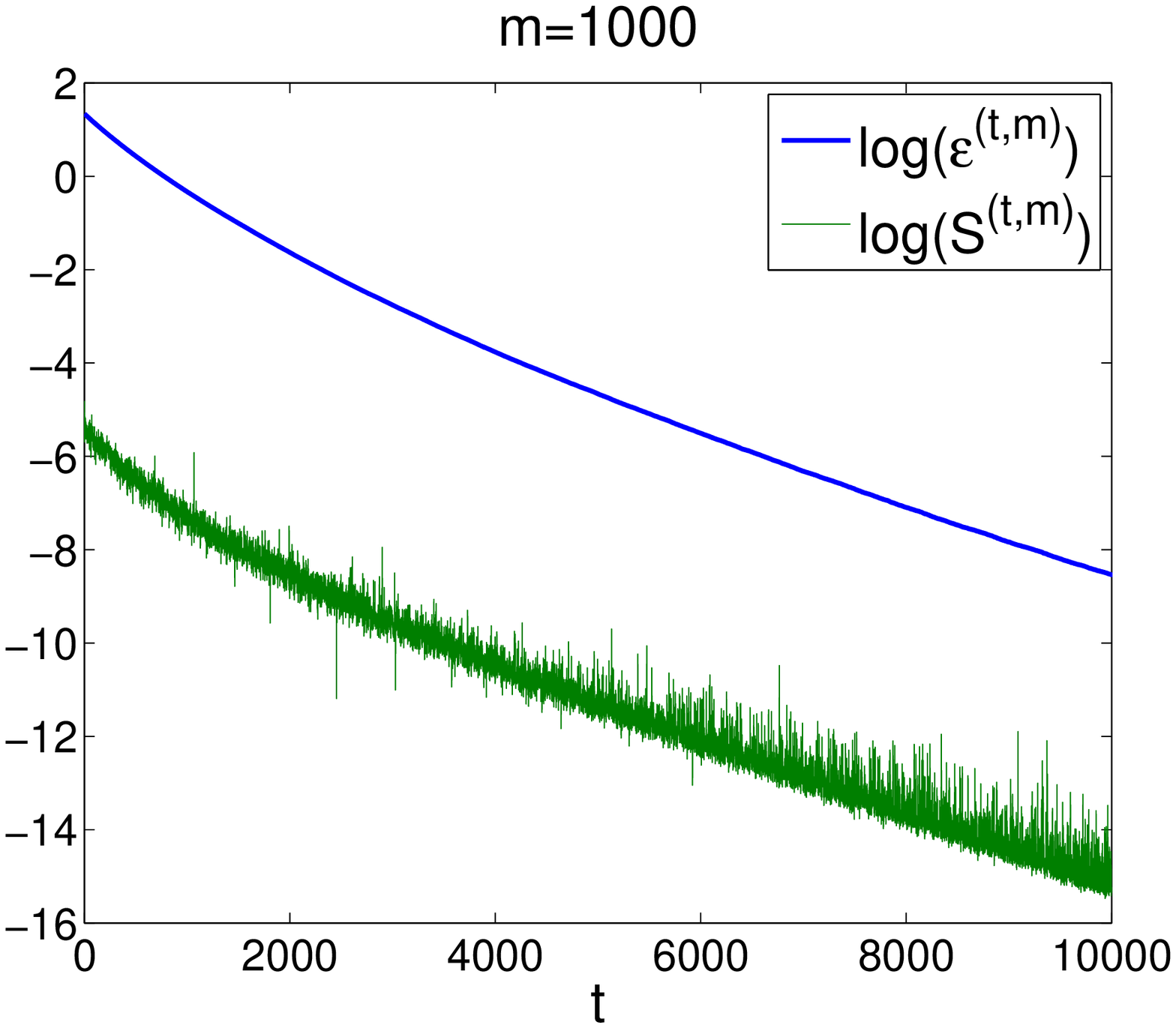}}}
\hspace*{-0.1in}\subfigure[varying $m$]{\includegraphics[scale=0.18]{{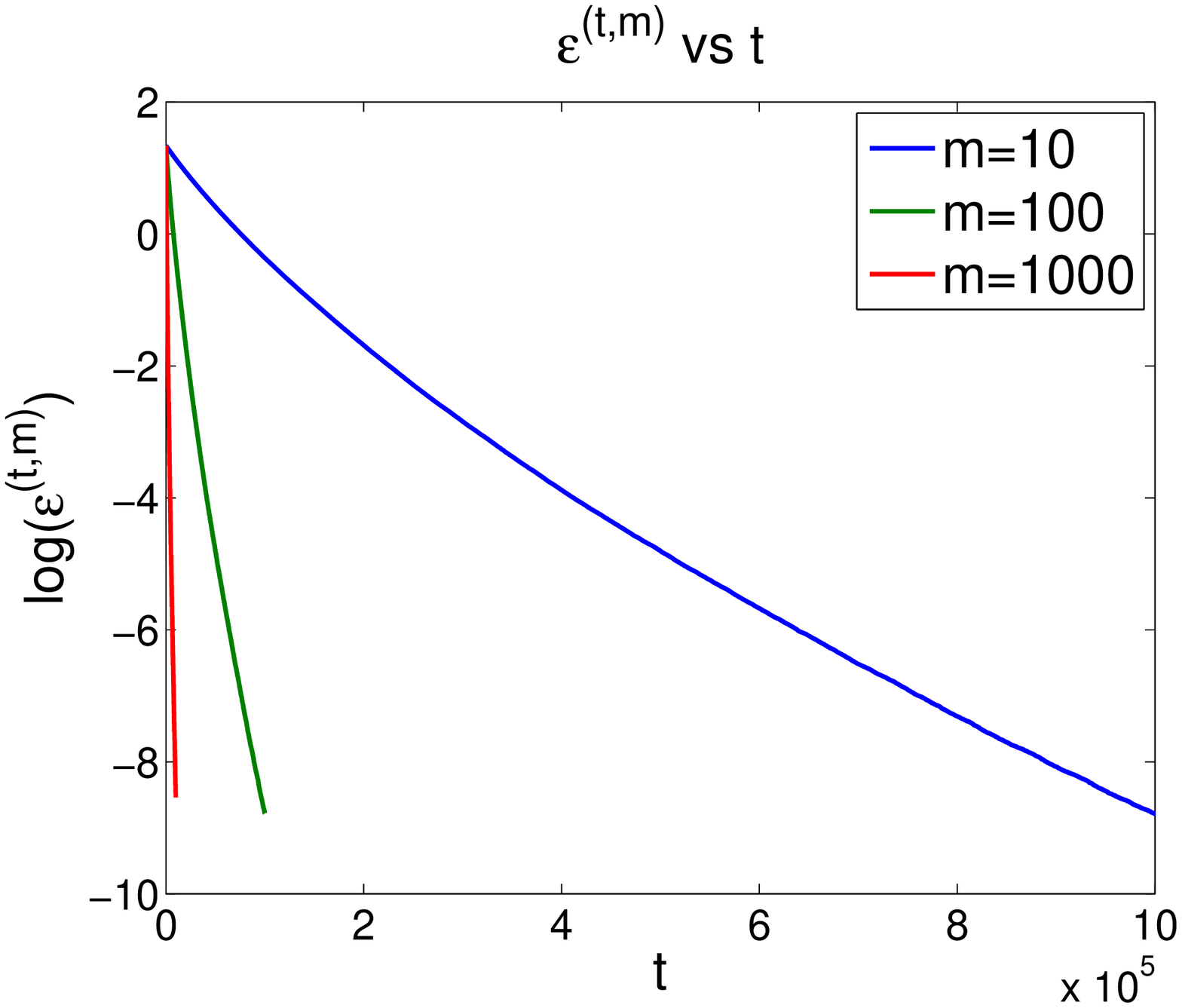}}}
\caption{Convergence of the practical DisDCA for varying $m$. The value of $\lambda$ is set to $10^{-5}$ and the data is randomly distributed to $K=5$ machines. The results in (a$\sim$d) are obtained on covtype data set using squared hinge loss for classification. The results in (e$\sim$h) are obtained on the synthetic data set using least square loss for regression.}\label{fig:pdisdca}
\end{figure}

 With fixed $m$, we have $\epsilon^{(t+1)} = \epsilon^{(t,m)}, \epsilon^{(t)} = \epsilon^{(t,0)}$ due to our definitions, then we have
  \begin{align*}
 \E[\epsilon^{(t+1)}] \leq \mu^m\E[\epsilon^{(t)}] +\E[S^{(t,m)}]
 \end{align*}
 
 Generally, bounding $S^{(t,m)}$ is a non-trivial task. Below, we provide some empirical studies to aid our analysis.  In Figure~\ref{fig:pdisdca}, we plot the curve for $S^{(t,m)}$ and also $\epsilon^{(t+1)}=\epsilon^{(t,m)}$ for fixed $m$.  The figures not only show that $S^{(m,t)}$ converges to zero, but also indicate that $S^{(t,m)}\leq \varepsilon(t, m)\epsilon^{(t+1)}$, where $\varepsilon(t, m)<1$  depends on $t, m$. Thus, we can establish 
\begin{align*}
(1 - \varepsilon(t,m))\E[\epsilon^{(t+1)}]\leq \left(1 - \frac{sK}{n}\right)^m\E[\epsilon^{(t)}]
\end{align*}

Then we can get 
\begin{align*}
\E[\epsilon^{(T)}]\leq \prod_{t=1}^T\left(\frac{1}{1-\varepsilon(t,m)}\right)\left(1 - \frac{sK}{n}\right)^{mT}\epsilon^{(0)}
\end{align*}
The dependence of the convergence on $t$ and $m$ is also verified by the results in Figure~\ref{fig:pdisdca}. At the earlier stages $\varepsilon(t,m)$ is smaller, therefore the slope of $\log(\epsilon^{(t,m)})$ is larger. Noting that $s = n/(cK+n)$, we have
\begin{align*}
\E[\epsilon^{(T)}]\leq \prod_{t=1}^T\left(\frac{1}{1-\varepsilon(t,m)}\right)\left(1 - \frac{1}{c + n/K}\right)^{mT}\epsilon^{(0)}
\end{align*}
We can also see that how the number of machines $K$ plays the role in the convergence rate. Let us consider an interesting case when $c= L/\lambda = O(n^{1/(1+\tau)})$, where $\tau\in(0,1]$~\citep{Yangnips13}. Then increasing $K$ upto $n^{\tau/(1+\tau)}$, the convergence rate can be speed-up via decreasing $\displaystyle \mu = 1 - \frac{1}{c + n/K}$. On the other hand, increasing $K$ would increase the communication cost. In practice, it is important to tune $m$ and $K$  to achieve the best  balance between computation and communication.


\section{Conclusions and Open Problems}\label{sec:4}
In this manuscript, we have made a progress in  analyzing  the practical DisDCA theoretically.  In particular, we have established the convergence rate of the practical DisDCA for orthogonal  data. We also provided an analysis of the practical DisDCA for genera data. Helped by empirical studies, we show that the practical DisDCA is able to achieve an exponential speed-up on the convergence by increasing the number of dual updates at each iteration. This result well justifies the superiority of the practical DisDCA over the naive variant, which only has a partially linear speed-up. 

There still exist open problems for future research. First, what is the analytical form of $\varepsilon(t,m)$ that upper bounds $S^{(t,m)}$ by $\epsilon^{(t,m)}$. Second, how to make the communication asynchronous and prove the convergence.

\bibliography{icml14}
\bibliographystyle{icml2014}
\section*{Appendix}
\subsection*{Proof of Lemma 1}
\begin{align*}
& A_k= -\phi_{k,i_j}^*(-\alpha_{k, i_j}^{t, j-1}-\Delta\alpha_{k, i_j}^{t, j}) - \frac{\lambda n}{2} \left\|w^{t, j-1}_k+\frac{1}{\lambda n}\Delta\alpha^{t,j}_{k, i_j}x_{k,i_j}\right\|^2_2 \\
&= -\phi_{k,i_j}^*(-\alpha_{k, i_j}^{t, j-1}-\Delta\alpha_{k, i_j}^{t, j})- \frac{\lambda n}{2} \|w^{t, j-1}_k\|_2^2 -{\Delta\alpha^{t,j}_{k, i_j}x_{k,i_j}^{\top}w_{k}^{t, j-1}} - \frac{1}{\lambda n}\|\Delta\alpha^{t,j}_{k, i_j}x_{k,i_j}\|^2_2\\
&= -\phi_{k,i_j}^*(-\alpha_{k, i_j}^{t, j-1}-\Delta\alpha_{k, i_j}^{t, j})- \frac{\lambda n}{2} \|w^{t, j-1}_k\|_2^2 -{\Delta\alpha^{t,j}_{k, i_j}x_{k,i_j}^{\top}(u_{k}^{t, j-1} - \bar w_{k}^{t-1})} - \frac{1}{\lambda n}\|\Delta\alpha^{t,j}_{k, i_j}x_{k,i_j}\|^2_2 \\
&= -\phi_{k,i_j}^*(-\alpha_{k, i_j}^{t, j-1}-\Delta\alpha_{k, i_j}^{t, j})- \frac{\lambda n}{2} \|w^{t, j-1}_k\|_2^2 -{\Delta\alpha^{t,j}_{k, i_j}x_{k,i_j}^{\top}u_{k}^{t, j-1}} - \frac{1}{\lambda n}\|\Delta\alpha^{t,j}_{k, i_j}x_{k,i_j}\|^2_2\\
&\geq-(1-s)\phi_{k,i_j}^*(-\alpha_{k, i_j}^{t, j-1})- s\phi^*_{k, i_j}(-\omega_{k, i_j}^{t, j-1})+ \frac{1}{2L}s(1-s)\|\omega_{k,i_j}^{t, j-1} - \alpha_{k,i_j}^{t, j-1}\|_2^2 \\
&-\frac{\lambda n}{2} \|w^{t, j-1}_k\|_2^2 -{s(\omega_{k,i_j}^{t, j-1}-\alpha^{t,j-1}_{k, i_j})x_{k,i_j}^{\top}w_{k}^{t, j-1}} - \frac{1}{\lambda n}s^2(\omega_{k,j_j}^{t, j-1}-\alpha^{t,j-1}_{k, i_j})\|x_{k,i_j}\|^2_2\\ 
&\geq - s(\phi_{k,i_j}^*(-\omega_{k,i_j}^{t,j-1})+\omega^{t,j-1}_{k,i_j}x_{k,i_j}^{\top}w^{t,j-1}_{k})\\
&+s\phi^*_{k,i_j}(-\alpha^{t,j-1}_{k,i_j}) ) + s\alpha_{k, i_j}^{t, j-1}x_{k,i_j}^{\top}w^{t, j-1}_{k} \underbrace{- \phi^*_{k,i_j}(-\alpha^{t,j-1}_{k,i_j})  -\frac{\lambda n}{2} \|w^{t, j-1}_k\|_2^2}\limits_{B_k} \\
& + \frac{s}{2}\left(\frac{(1-s)}{L} - \frac{s}{\lambda n}\|x_{k,i_j}\|_2^2\right)\|\omega_{k,j_j}^{t, j-1}-\alpha^{t,j-1}_{k, i_j}\|_2^2\\
&\geq s\left[\phi_{k,i_j}(x_{k,i_j}^{\top}w_k^{t,j-1})+\phi^*_{k,i_j}(-\alpha^{t,j-1}_{k,i_j})+\alpha_{k, i_j}^{t, j-1}x_{k,i_j}^{\top}w^{t, j-1}_{k}\right]
 \end{align*}
By setting $s = n/(c + n)$, we have
 \begin{align*}
 &n(D(\alpha^{t,j}) - D(\alpha^{t,j-1})) \\
 &\geq \sum_{k=1}^Ks\left[\phi_{k,i_j}(x_{k,i_j}^{\top}w_k^{t,j-1})+\phi^*_{k,i_j}(-\alpha^{t,j-1}_{k,i_j})+\alpha_{k, i_j}^{t, j-1}x_{k,i_j}^{\top}w^{t, j-1}_{k}\right]\\
 &=\sum_{k=1}^Ks\left[\phi_{k,i_j}(x_{k,i_j}^{\top}w^{t,j-1})+\phi^*_{k,i_j}(-\alpha^{t,j-1}_{k,i_j})+\alpha_{k, i_j}^{t, j-1}x_{k,i_j}^{\top}w^{t, j-1}\right]
  \end{align*}
 The  Lemma~\ref{lem:2} can be proved by showing that the expectation of the R.H.S of the above inequality is equal to the duality gap $P(w^{t,j-1}) - D(\alpha^{t,j-1})$ with appropriate scaling.

 \subsection*{Proof of Lemma~\ref{lem:3}}
 By the definition of $D(\alpha)$, we have 
\begin{align*}
&n[D(\alpha^{t, j}) - D(\alpha^{t, j-1})] \\
&= \left[\sum_{k}\sum_{i}-\phi^*_{k,i}(-\alpha_{k,i}^{t,j}) - \frac{\lambda n}{2}\|w^{t, j}\|_2^2\right] -  \left[\sum_{k}\sum_{i}-\phi^*_{k,i}(-\alpha_{k,i}^{t,j-1}) - \frac{\lambda n}{2}\|w^{t, j-1}\|_2^2\right]\\
&\geq  \sum_{k=1}^K- \phi^*_{k,i_j}(-\alpha^{t, j}_{k,i_j}) - \frac{\lambda n}{2}\left\|w^{t,j-1} + \sum_{k=1}^K\frac{1}{\lambda n}\Delta\alpha^{t, j}_{k, i_j}x_{k,i_j}\right\|_2^2-\sum_{k=1}^K - \phi^*_{k,i_j}(-\alpha^{t, j-1}_{k,i_j}) - \frac{\lambda n}{2}\|w^{t,j-1}\|_2^2\\
&\geq  \sum_{k=1}^K- \phi^*_{k,i_j}(-\alpha^{t, j}_{k,i_j}) - \frac{\lambda n}{2}\left\|w^{t,j-1}\right\|_2^2 -  \sum_{k=1}^K\Delta\alpha^{t, j}_{k, i_j}x_{k,i_j}^{\top}w^{t, j-1} - \frac{1}{2\lambda n}\left\|\sum_{k=1}^K\Delta\alpha^{t, j}_{k, i_j}x_{k,i_j}\right\|_2^2\\
&-\sum_{k=1}^K - \phi^*_{k,i_j}(-\alpha^{t, j-1}_{k,i_j}) - \frac{\lambda n}{2}\|w^{t,j-1}\|_2^2\\
&\geq  \sum_{k=1}^K- \phi^*_{k,i_j}(-\alpha^{t, j}_{k,i_j}) - \frac{\lambda n}{2}\left\|w^{t,j-1}\right\|_2^2\\
& -  \sum_{k=1}^K\Delta\alpha^{t, j}_{k, i_j}x_{k,i_j}^{\top}w^{t, j-1} - \frac{1}{2\lambda n_k}\sum_{k=1}^K\left\|\Delta\alpha^{t, j}_{k, i_j}x_{k,i_j}\right\|_2^2-\sum_{k=1}^K - \phi^*_{k,i_j}(-\alpha^{t, j-1}_{k,i_j}) - \frac{\lambda n}{2}\|w^{t,j-1}\|_2^2\\
&\geq  \sum_{k=1}^K- \phi^*_{k,i_j}(-\alpha^{t, j-1}_{k,i_j}-\Delta\alpha^{t,j}_{k, i_j})  -  \sum_{k=1}^K\Delta\alpha^{t, j}_{k, i_j}x_{k,i_j}^{\top}u_k^{t, j-1} - \frac{K}{2\lambda n}\sum_{k=1}^K\left\|\Delta\alpha^{t, j}_{k, i_j}x_{k,i_j}\right\|_2^2\\
&-\sum_{k=1}^K - \phi^*_{k,i_j}(-\alpha^{t, j-1}_{k,i_j})  + \sum_{k=1}^K\Delta\alpha^{t, j}_{k, i_j}x_{k,i_j}^{\top}(u_k^{t, j-1}  - w^{t,j-1}) \\
&\geq  \sum_{k=1}^K- (1-s)\phi^*_{k,i_j}(-\alpha^{t, j-1}_{k,i_j})-s\phi^*_{k,i_j}(-\omega^{t,j-1}_{k, i_j}) + \frac{1}{2L} s(1-s)\|\omega^{t,j-1}_{k, i_j}-\alpha^{t,j-1}_{k, i_j})\|_2^2 \\
&-  \sum_{k=1}^Ks(\omega^{t,j-1}_{k, i_j}-\alpha^{t,j-1}_{k, i_j})x_{k,i_j}^{\top}u_k^{t, j-1} - \frac{K}{2\lambda n}\sum_{k=1}^Ks^2\left(\omega^{t,j-1}_{k, i_j}-\alpha^{t,j-1}_{k, i_j})^2\|x_{k,i_j}\right\|_2^2\\
&-\sum_{k=1}^K - \phi^*_{k,i_j}(-\alpha^{t, j-1}_{k,i_j})  + \sum_{k=1}^K\Delta\alpha^{t, j}_{k, i_j}x_{k,i_j}^{\top}(u_k^{t, j-1}  - w^{t,j-1})\\
&\geq  \sum_{k=1}^Ks\phi^*_{k,i_j}(-\alpha^{t, j-1}_{k,i_j})+  \sum_{k=1}^Ks\alpha^{t,j-1}_{k,i_j}x_{k,i_j}^{\top}u^{t,j-1}_{k}+[-s\phi^*_{k,i_j}(-\omega^{t,j-1}_{k, i_j})  - s\omega^{t,j-1}_{k,i_j}x_{k,i_j}^{\top}u^{t,j-1}_{k}]\\
&+ s\sum_{k=1}^K\left(\frac{1}{2L} (1-s) - \frac{1}{2\lambda n_k}\|x_{k,i_j}\|_2^2\right)\|\omega^{t,j-1}_{k, i_j}-\alpha^{t,j-1}_{k, i_j}\|_2^2  + \sum_{k=1}^K\Delta\alpha^{t, j}_{k, i_j}x_{k,i_j}^{\top}(u_k^{t, j-1}  - w^{t,j-1}) \\
&\geq  \sum_{k=1}^Ks\phi^*_{k,i_j}(-\alpha^{t, j-1}_{k,i_j})+  \sum_{k=1}^Ks\alpha^{t,j-1}_{k,i_j}x_{k,i_j}^{\top}w^{t,j-1}+[-s\phi^*_{k,i_j}(-\omega^{t,j-1}_{k, i_j})  - s\omega^{t,j-1}_{k,i_j}x_{k,i_j}^{\top}w^{t,j-1}]\\
&+ s\sum_{k=1}^K\left(\frac{1}{2L} (1-s) - \frac{1}{2\lambda n_k}\|x_{k,i_j}\|_2^2\right)\|\omega^{t,j-1}_{k, i_j}-\alpha^{t,j-1}_{k, i_j}\|_2^2  + \sum_{k=1}^K(\Delta\alpha^{t, j}_{k, i_j} - s(\omega^{t,j-1}_{k, i_j}-\alpha^{t,j-1}_{k, i_j}))x_{k,i_j}^{\top}(u_k^{t, j-1}  - w^{t,j-1}) \\
&\geq s\left[ \sum_{k=1}^K\phi^*_{k,i_j}(-\alpha^{t, j-1}_{k,i_j})+\phi_{k,i_j}(x_{k,i_j}^{\top}w^{t,j-1})+  \sum_{k=1}^K\alpha^{t,j-1}_{k,i_j}x_{k,i_j}^{\top}w^{t,j-1}\right] \\
&+ s\sum_{k=1}^K\left(\frac{1}{2L} (1-s) - \frac{1}{2\lambda n_k}\|x_{k,i_j}\|_2^2\right)\|\omega^{t,j-1}_{k, i_j}-\alpha^{t,j-1}_{k, i_j}\|_2^2  \\
&-  \underbrace{\color{blue}\sum_{k=1}^K(s(\omega^{t,j-1}_{k, i_j}-\alpha^{t,j-1}_{k, i_j})- \Delta\alpha^{t, j}_{k, i_j} )x_{k,i_j}^{\top}(u_k^{t, j-1}  - w^{t,j-1})}\limits_{nR^{t,j}}
\end{align*}

\end{document}